\documentclass[usenatbib, useAMS]{mnras}

\usepackage{latexsym,amssymb}
\usepackage[fleqn]{amsmath}
\usepackage{graphics,graphicx}
\usepackage[absolute]{textpos}
\usepackage{natbib}
\usepackage[T1]{fontenc}
\usepackage{rotating}
\usepackage{aecompl}
\usepackage{caption}
\usepackage{adjustbox,lipsum}
\usepackage{array}
\usepackage{dcolumn}

\usepackage{amsmath} 
\usepackage{enumitem}
\usepackage{lscape}
\usepackage{enumitem}

\newcommand{\Fe}{Fe$_{5015}$}
\newcommand{\Mg}{Mg$_b$}

\newcommand{\Fee}{Fe$_{5015, {\rm e}}$}
\newcommand{\Mge}{Mg$_{b, {\rm e}}$}

\title{On the observational diagnostics to separate classical and disk-like bulges}

\author[L. Costantin et al.]
{Luca Costantin,$^{1}$\thanks{luca.costantin@studenti.unipd.it}
E.~M.~Corsini,$^{1, 2}$
J.~M\'endez-Abreu,$^{3, 4}$
L.~Morelli,$^{1, 2, 5}$
\newauthor E.~Dalla Bont\`a,$^{1, 2}$
and A.~Pizzella$^{1, 2}$
\\
\\
$^{1}$Dipartimento di Fisica e Astronomia `G. Galilei',
  Universit\`a di Padova, vicolo dell'Osservatorio 3, I-35122 Padova,
  Italy\\
$^{2}$INAF - Osservatorio Astronomico di Padova, vicolo
dell'Osservatorio 5, I-35122 Padova, Italy\\
$^{3}$Instituto de Astrof\'isica de Canarias, Calle V\'ia L\'actea s/n,
E-38200 La Laguna, Tenerife, Spain\\
$^{4}$Departamento de Astrof\'isica, Universidad de La Laguna, Calle
Astrof\'isico Francisco S\'anchez s/n, E-38205 La Laguna, Tenerife,
Spain\\
$^{5}$Instituto de Astronomía y Ciencias Planetarias, Universidad de Atacama, Copiapó, Chile}

\begin{document}

\date{\it Accepted version on 2018 June 27}
\pubyear{2018}

\label{firstpage}
\pagerange{\pageref{firstpage}--\pageref{lastpage}}
\maketitle

\begin{abstract}
Flattened bulges with disk-like properties are considered to be the
end product of secular evolution processes at work in the inner
regions of galaxies. On the contrary, classical bulges are
characterized by rounder shapes and thought to be similar to
low-luminosity elliptical galaxies.
We aim at testing the variety of observational diagnostics which are
commonly adopted to separate classical from disk-like bulges in
nearby galaxies.
We select a sample of eight unbarred lenticular galaxies to be
morphologically and kinematically undisturbed with no evidence of
other components than bulge and disk. We analyze archival data of
broad-band imaging from SDSS and
integral-field spectroscopy from the ATLAS$^{\rm 3D}$ survey to
derive the photometric and kinematic properties, line-strength
indices, and intrinsic shape of the sample bulges.
We argue that the bulge S\'ersic index is a poor diagnostics to
discriminate different bulge types. We find that the
combination of line-strength with either kinematic 
or photometric diagnostics does not provide a
clear separation for half of the sample bulges. 
We include for the first time
the intrinsic three-dimensional shape of bulges as
a possible discriminant of their nature.
All bulges turn out to be thick
oblate spheroids, but only one has a flattening consistent with
that expected for outer disks.
We conclude that bulge classification may be difficult even adopting
all observational diagnostics proposed so far and that classical and
disk-like bulges could be more confidently identified by considering
their intrinsic shape.
\end{abstract}




\section{Introduction}
\label{sec:introduction}

The term \emph{bulge} entered the vocabulary of extragalactic
astrophysics in the early 1940s when Edwin Hubble, debating about the
direction of rotation of spiral galaxies, denominated their central
protuberances as nuclear bulges \citep{Hubble1943}. This word was
beyond doubt referred to an observed structure sticking out from the
galactic plane of lenticular and spiral galaxies. Later the term also assumed
a more physical meaning related to the processes driving the formation
and evolution of galaxies. For example, \citet{Renzini1999} looked at
bulges as elliptical galaxies surrounded by an outer disk or,
conversely, considered elliptical galaxies as bulges that somehow were
not able to acquire and preserve a disk component.

Several operative definitions were proposed to outline the bulge
component by analyzing the surface brightness distribution of the
host galaxy.
\citet{Kent1986} pointed out that the galaxy isophotes in the bulge
region are more round than those measured in the disk region, assuming
that both bulge and disk have elliptical isophotes of constant but
different axial ratios; but this diagnostics does not work for
axisymmetric bulges seen face-on.
\citet{Carollo1999} defined bulges as the inner components of galaxies
whose light contribution exceeds the inward extrapolation of the
exponential disk, with the advantage of dealing with all disk galaxies regardless of their
inclination.  
The recent generalization of multi-component photometric
decompositions have led to define bulges as the central brightest
component following a S\'ersic profile \citep{Mendez2017}.

The current picture divides galactic bulges into two broad classes,
namely the \emph{classical bulges} and \emph{disk-like bulges},
according to their different formation processes
\citep{Athanassoula2005}. Classical bulges form via
dissipative collapse of protogalactic gas clouds \citep{Eggen1962} or grow
out of disk material triggered by satellite accretion during galaxy mergers
\citep{Aguerri2001, Hopkins2009}.  
But, they could also grow by the coalescence of giant clumps in primordial disks
\citep{Bournaud2016}.
Thus, these systems are expected to 
present hot orbits for their stars, 
having shapes that resemble elliptical galaxies and
being intrinsically less flat than their surrounding disks.
By contrast, disk-like bulges are thought to be the product of secular processes driven by internal
processes in galaxies, responsible to rearrange 
energy and mass in their center
\citep{Kormendy2004, Kormendy2016}.  
Due to their disky nature, these bulges are
expected to preserve the intrinsic flatness of the original disk as well as
its cold orbits.
We choose to refer to disk-like bulges rather than to
\emph{pseudobulges} to avoid misinterpretations. 
\citet{Kormendy1993} introduced the notion of \emph{bulge-like disks}
in the framework of a relaxed universe, when
slow secular processes drive the evolution of galaxies rather than
mergers. A decade later, \citet{Kormendy2004} renamed these objects as pseudobulges.
Nowadays the term pseudobulge is adopted to describe a
wide assortment of bulge types ranging from boxy/peanut-shaped structures,
which are actually bars seen from particular viewing angles
\citep[e.g.,][]{Laurikainen2005, Mendez2008b, Erwin2017}, to those characterized by the presence of inner
components, like ovals, nuclear bars or disks, rings, dust lanes, and
spiral arms \citep{Fisher2010}, 
to bulges with a fainter surface
brightness compared to elliptical galaxies of the same size
\citep{Gadotti2009}. 

A proper classification of bulges based 
on their formation scenarios is highly desirable, 
although yet premature. Nevertheless, given the 
variety of formation processes, different photometric,
kinematic, and stellar population properties are thought to
differentiate different formation paths in bulges.
Recently, \citet{Fisher2016} have extended the early classification scheme by
\citet{Kormendy2004} by reviewing the bulge properties, providing an
updated list of observational criteria to classify classical and
disk-like bulges, and addressing their demography in nearby galaxies.
In this paper, we homogeneously apply all these criteria to a
well-defined sample of bulges to investigate which one, or which
combination of them, is more effective characterizing classical and
disk-like bulges.
Indeed, one or few of the criteria given by
\citet{Fisher2016} are commonly adopted to classify bulges 
\citep[e.g.,][]{Lorenzo2014, Vaghmare2015, Mishra2017}. 
However, even though this view divides bulges into 
well-defined classes, the complexity of galaxy formation suggests a continuity
rather than a bimodality of bulge properties. 
Therefore, this picture could be reframed in term of the bulge intrinsic shape,
where bulges are sorted from more to less flattened systems, as well as elliptical
galaxies are classified on the Hubble tuning fork. In this context, boxy/peanut structures 
represent a separate or a related case of interest.
Furthermore, the coexistence of composite bulges in
the same galaxy leads to a more complicated picture and poses more
challenges in galaxy formation mechanisms \citep{Mendez2014, Erwin2015}.
Thus, to preserve the original distinction of bulges according to
their formation mechanisms, the bulge intrinsic shape could
be useful to unveil the information on bulge orbits and evolutionary processes,
from classical to disk-like systems. 
Moreover, we also propose to characterize the bulge intrinsic shape,
which it is a diagnostics not yet adopted in bulge classification,
in order to study its actual interplay with other bulge diagnostics.
To address this issue, we select our
sample bulges from the volume-limited ATLAS$^{\rm 3D}$ survey of
early-type galaxies \citep[ETGs,][]{Cappellari2011}. It opened a new
era for the integral-field spectroscopic surveys of nearby galaxies by
solving some of the long-standing puzzles about the kinematic
peculiarities \citep{Krajnovic2011}, dynamical status
\citep{Emsellem2011}, and stellar populations \citep{McDermid2015} of
ETGs \citep[see also][for a review]{Cappellari2016}. On the other
hand, we benefit from the ATLAS$^{\rm 3D}$ legacy of two-dimensional
maps of ionized-gas \citep{Sarzi2013}, stellar kinematics
\citep{Emsellem2011}, and line-strength indices \citep{Scott2013} to
accurately measure the properties of our sample bulges in a consistent
way.

The paper is organized as follows. We list the observational criteria
for classifying bulges in Sect.~\ref{sec:criteria}. We present the
galaxy sample in Sect.~\ref{sec:sample}. We analyze the surface
brightness distribution of the sample galaxies in
Sect.~\ref{sec:photometry}.  We recover the intrinsic shape of the
sample bulges in Sect.~\ref{sec:shape}.  We analyze the stellar
kinematics and line-strength indices of the sample bulges in
Sect.~\ref{sec:spectroscopy}. We investigate whether our bulges follow
the same scaling relations traced by elliptical galaxies and large
bulges in Sect.~\ref{sec:scaling}. We discuss the classification of
the sample bulges in Sect.~\ref{sec:discussion}. We present our
conclusions in Sect.~\ref{sec:conclusions}.


\section{Observational criteria for bulge classification}
\label{sec:criteria}

In their review about the properties and classification of bulges in
nearby galaxies, \citet{Fisher2016} proposed a number of observational
criteria (i.e., category I diagnostics) that allow to mark a
relatively clean separation between classical and disk-like bulges and
to statistically classify all the bulges within a sample. They also
identified other observational properties (i.e., category II
diagnostics) that can be used to classify single bulges, but can not
be applied to the whole bulge population. Finally, they gave a few
additional criteria (i.e., category III diagnostics), which are
supposed to be necessary (but not sufficient) to identify a bulge as
classical. Here, we provide a summary of the
observational criteria given by \citet{Fisher2016} grouped according
to their category. 

Classical bulges are thought to:

\begin{itemize}[leftmargin=1.3cm]
\item[(I-1)$_{\rm C}$] show no spiral or ring structures in the region
  where they dominate the galaxy light, as shown by optical images
  taken at high spatial resolution (${\emph FWHM}_{\rm PSF} < 100$ pc);
\item[(I-2)$_{\rm C}$] have S\'ersic index $n > 2$;
\item[(I-3)$_{\rm C}$] show \Mg-- $\sigma$ and \Mg--\Fe\  correlations 
  consistent with those of elliptical galaxies; 
\item[(I-4)$_{\rm C}$] have a strongly peaked radial profile of
  stellar velocity dispersion $\sigma$, with a gradient d$\log(\sigma)$/d$\log(r) < -
  0.1$ within [$r_{\rm min}$, $r_{25\%}$], where $r_{\rm min} = {\emph
    FWHM}_{\rm PSF}$ to exclude the inner regions of the kinematic
  maps which are most affected by blurring effects of the point
  spread function (PSF) and $r_{25\%}$ is the radius where the surface
  brightness contribution of the bulge exceeds that of the disk by
  25\%;
\item[(II-1)$_{\rm C}$] have central velocity dispersion $\sigma_0 >
  130$ km s$^{-1}$;
\item[(III-1)$_{\rm C}$] be consistent with the fundamental plane
  relation (FPR) of elliptical galaxies;
\item[(III-2)$_{\rm C}$] show low specific star formation rate $sSFR <
  10^{-11}$ yr$^{-1}$ \citep[but this is not applicable to lenticular
    galaxies;][]{Kormendy2016};
\item[(III-3)$_{\rm C}$] rarely present extremely blue colors (e.g., $B - V <
  0.65$). 
\end{itemize}

Disk-like bulges are supposed to:

\begin{itemize}[leftmargin=1.3cm]
\item[(I-1)$_{\rm D}$] show spiral or ring structures in the region
  where they dominate the galaxy light, as shown by optical images
  taken at high spatial resolution (${\emph FWHM}_{\rm PSF} < 100$ pc);
\item[(I-2)$_{\rm D}$] have S\'ersic index $n < 2$;
\item[(I-3)$_{\rm D}$] show line-strength offset $\Delta$\Mg$ < 0.7$
  \AA\ compared to the \Mg-- $\sigma$ correlation, or $\Delta$\Mg$ <
  0.7$ \AA\ compared to the \Mg--\Fe\ relation of elliptical galaxies;
\item[(I-4)$_{\rm D}$] present a stellar velocity dispersion radial
  profile that satisfies d$\log(\sigma)$/d$\log(r) > - 0.1$ or
  $\langle v^2\rangle/\langle \sigma^2 \rangle \ge 0.35$ within
  [$r_{\rm min}$, $r_{25\%}$];
\item[(II-1)$_{\rm D}$] be low surface-brightness outliers in the
  Kormendy relation (KR) of elliptical galaxies;
  \item[(II-2)$_{\rm D}$] present high specific star formation rate
    $sSFR > 10^{-11}$ yr$^{-1}$ \citep[but this is not applicable to
    lenticular galaxies;][]{Kormendy2016};
\item[(II-3)$_{\rm D}$] have line-strength indices \Fe$ <
  3.95$ \AA\ and \Mg$ < 2.35$ \AA;
\item[(II-4)$_{\rm D}$] be low-$\sigma$ outliers in the Faber-Jackson
  relation (FJR) of elliptical galaxies;
\item[(II-5)$_{\rm D}$] show blue optical colors (e.g., $B - V <
  0.5$).
\end{itemize}

It is worth noting that, even if the different observational
properties of bulges can be explained in terms of formation process
and evolutionary history, most of the above criteria are based on an
\emph{a priori} separation between classical and disk-like bulges
which is usually done with a visual morphological classification.  For
these reasons, different authors \citep[e.g.,][]{Graham2008,
  Mendez2018} challenged these criteria by pointing out that they can
easily lead to misclassification  when one or few
of them are adopted to select a particular type of bulge.
  

\section{Sample selection}
\label{sec:sample}

\begin{table}
\caption{Galaxies rejected after the photometric decomposition.}
\centering
\begin{tabular}{cccc}
\hline
\hline
Galaxy & Motivation & Galaxy & Motivation \\
(1)    & (2)     & (3)    &  (4)   \\
\hline
IC~3631	 	& 	morph. 	&	NGC~4474	&	inc.			\\
NGC~525	 	&	inc.		&	NGC~4638	&	inc.			\\
NGC~2577	&	inc.	 	&	NGC~5273	&	morph. 		\\
NGC~2685	&	inc.		&	NGC~5485	&	ell.			\\
NGC~3648	&	morph. 	&	NGC~6278	&	morph. 		\\
NGC~3665	&	ell.		&	PGC~35754	&	ell.			\\
NGC~4379	&	morph. 	&	UGC~9519	&	inc.	 		\\
\hline
\end{tabular}
\begin{minipage}{8.5cm} 
{\bf Notes.\/} Columns (1), (3): galaxy name. Columns (2), (4): motivation for rejecting the galaxy; ell. = elliptical galaxy,
morph. = bar and/or spiral arms, and incl. = too highly inclined galaxy.
\end{minipage}
\label{tab:rejected}
\end{table}

\begin{table*}
\centering
\caption{Properties of the sample galaxies.}
\begin{tabular}{ccccccccc}
\hline
Galaxy			&	 RA & DEC 		&	$d$ 		& scale  &	$m_i$		&	$M_i$ &  $R_{\rm e, gal}$ & $\theta_{\epsilon}$      \\
				&	 {[$^{\rm h}\;^{\rm m}\;^{\rm s}$]} & {[$^\circ\;'\;''$]} 	&	{[Mpc]} 	& [pc arcsec$^{-1}$] & 	{[mag]}	 	&	 {[mag]} & [arcsec]	& {[$^{\circ}$]}	   \\
(1)				&	(2)		&	(3)			&	(4)		& (5) 	& (6)			& (7) & (8) & (9)\\
\hline
NGC~3156		&	10 12 41.25	&	$+03$ 07 45.69	&	21.8	&	106	&	12.05		&	$-19.64$		&	17.4	&	60	\\
NGC~3245		&	10 27 18.39	&	$+28$ 30 26.79	&	20.3	&	98	&	10.39		&	$-21.15$		&	25.1	&	57	\\
NGC~3998		&	11 57 56.13	&	$+55$ 27 12.92	&	13.7	&	66 	&	11.04		&	$-19.64$		&	20.0	&	39	\\
NGC~4578		&	12 37 30.56	&	$+09$ 33 18.25	&	16.3	&	79 	&	11.07		&	$-20.00$		&	32.4	&	45	\\
NGC~4690		&	12 47 55.52	&	$-01$ 39 21.83	&	40.2	&	195	&	12.26		&	$-20.76$		&	17.8	&	45	\\
NGC~5687		&	14 34 52.40	&	$+54$ 28 33.05	&	27.2	&	131	&	11.71		&	$-20.46$		&	22.9	&	51	\\
NGC~6149		&	16 27 24.23	&	$+19$ 35 49.91	&	37.2	&	180	&	12.65		&	$-20.20$		&	10.7	&	47	\\
NGC~7457		&	23 00 59.93	&	$+30$ 08 41.79	&	12.9	&	63	&	10.76		&	$-19.79$	  	& 	36.3	&	58  	\\
\hline
\end{tabular}
\begin{minipage}{180mm} 
{\bf Notes.\/} Column (1): galaxy name. Columns (2), (3): right ascension and declination
(J2000.0).  Column (4): galaxy distance from \citet{Cappellari2011}. Column (5):
conversion factor from arcsec to parsec. Column (6):
composite-model apparent $i$-band magnitude (cmodel) of the galaxy
from SDSS. Column (7): absolute $i$-band magnitude of the galaxy. Column (8): 
circularized effective radius of the galaxy from \citet{Cappellari2011}.
Column (9): galaxy inclination $\theta_{\epsilon} = \arccos(1 - \epsilon)$, where
$\epsilon$ is galaxy ellipticity at $R_{\rm e, gal}$ from
\citet{Krajnovic2011}.
\end{minipage}
\label{tab:sample}
\end{table*}

We selected our sample of unbarred lenticular galaxies among the
nearby ETGs of the ATLAS$^{\rm 3D}$ survey \citep{Cappellari2011}.
Since this pilot project aims to understand the interplay of different bulge
diagnostics, the sample galaxies were selected having in mind the
simplest systems in terms of their structure, morphology, photometric
and kinematic properties, that is, lenticular galaxies. ATLAS$^{\rm 3D}$
provided the ideal starting point to challenge all the observational
diagnostics proposed so far, where recent surveys like 
Calar Alto Legacy Integral Field Area (CALIFA)
lack in spatial resolution to perform an exhaustive analysis
for later Hubble type galaxies.

First, we considered the 111 galaxies classified as unbarred
lenticular galaxies combining the information from \citet[][Hubble
  stage]{Cappellari2011} and \citet[][barredness]{Krajnovic2011},
since they are supposed to be the disk galaxies with the simplest
structure having only a bulge and a disk component. Then, we examined
only the 58 galaxies without any morphological or kinematic
peculiarity.  Indeed, we rejected all the galaxies with signatures of
interaction or merging, as it results from the visual inspection of
their Sloan Digital Sky Survey (SDSS) images and those with a
kinematically distinct cores or counter-rotating components
\citep{McDermid2006, Krajnovic2011}. Finally, we restricted our
analysis to the 22 galaxies with an inclination $\theta_{\epsilon} =
\arccos(1 - \epsilon) = [30^{\circ}, 60^{\circ}$], derived from the
axial ratio calculated from the global ellipticity measured within
$\sim3$ effective radii of the galaxy in \citet{Krajnovic2011} and
assuming an infinitesimally thin disk. This is required to perform a
successful photometric decomposition of the galaxy images.  This
sample was further curbed after performing the photometric
decomposition of the SDSS images (see Table~\ref{tab:rejected} 
and Sect.~\ref{sec:photometry} for
details). We found that (i) three galaxies turned out to be
ellipticals rather than lenticulars, (ii) five galaxies had a bar
and/or spiral arms, and (iii) six galaxies were too much inclined.

The final sample is composed of eight galaxies, for which we report the
main properties in Table~\ref{tab:sample}. 


\section{Surface photometry} 
\label{sec:photometry}

\subsection{Sloan Digital Sky Survey imaging}

We retrieved the \emph{i}-band images of the sample galaxies from the
Data Archive Server (DAS) of the SDSS Data Release 9 \citep{Ahn2012}.
All the archive images were already bias subtracted, flat-field
corrected, sky subtracted, and flux calibrated according to the
associated calibration information stored in the DAS.

We made use of the procedure described in \citet{Pagotto2017} to
measure the level of the sky background and its standard deviation
(Table \ref{tab:photometry}) after masking foreground stars, nearby
and background galaxies, residual cosmic rays, and bad pixels. We
found that our estimates of the sky background are systematically
lower by $0.3\%$ than those given by SDSS and we 
applied such a correction to the images. We trimmed the
sky-subtracted images to reduce the computing time when performing the
photometric decomposition. We centered each galaxy in a field of view
of at least $300\times300$ pixels$^2$ corresponding to $120\times120$
arcsec$^2$. Finally, we modeled the PSF of the resulting images with a
circular Moffat function \citep{Moffat1969} with the shape parameters
measured directly from the field stars (Table \ref{tab:photometry}).

We fitted elliptical isophotes to the galaxy images with the
\texttt{ellipse} task in IRAF\footnote{Image Reduction and Analysis
Facility is distributed by the National Optical Astronomy
Observatory (NOAO), which is operated by the Association of
Universities for Research in Astronomy (AURA), Inc. under
cooperative agreement with the National Science Foundation.}
\citep{Jedrzejewski1987} after masking out as much as possible dust
patches and lanes. We thus derived the radial
profiles of azimuthally-averaged surface brightness $\mu$, ellipticity
$\epsilon$, and position angle $PA$ to be used in the two-dimensional
photometric decomposition to estimate the starting guesses of the
galaxy structural parameters.

\begin{table}
\caption{Characteristics of the \emph{i}-band SDSS images of the sample galaxies.}
\label{tab:photometry}
\centering
\begin{tabular}{ccccccc}
\hline
Galaxy & Gain                & RON      & Sky   & $\emph{FWHM}$     & $\beta$\\
       & [$e^{-}$~ADU$^{-1}$] & [$e^{-}$]       & [ADU] & [arcsec] &        \\
(1)    & (2)                 & (3)      & (4)   & (5)      & (6)    \\
\hline
NGC~3156	 	&	5.2	&	14.5	&	144 $\pm$ 5	&	1.3		&	3.2	\\
NGC~3245	 	&	6.6	&	16.4	&	186 $\pm$ 6	&	1.3		&	4.6	\\
NGC~3998	 	&	4.6	&	13.0	&	177 $\pm$ 5	&	1.1		&	4.1	\\
NGC~4578		&	4.9	&	10.4	&	160 $\pm$ 5	&	1.1		&	4.2	\\
NGC~4690		&	6.6	&	16.4	&	206 $\pm$ 5	&	1.0		&	3.6	\\
NGC~5687		&	6.6	&	16.4	&	224 $\pm$ 5	&	1.2		&	3.5	\\
NGC~6149		&	4.9	&	10.4	&	124 $\pm$ 4	&	1.1		&	4.5	\\
NGC~7457		&	6.6	&	16.4	&	211 $\pm$ 6	&	0.9		&	6.7	\\
\hline
\end{tabular}
\begin{minipage}{8.5cm} 
{\bf Notes.\/} Column (1) galaxy name. Columns (2), (3): image gain and read-out provided
by SDSS. Column (4): measured sky level and corresponding standard
deviation. Columns (5), (6): $\emph{FWHM}$ and $\beta$ parameter measured for the
circular Moffat PSF.
\end{minipage}
\end{table}

\subsection{Photometric decomposition}

We performed the two-dimensional photometric decomposition of the SDSS
images of the 22 unbarred lenticular galaxies taken from the
ATLAS$^{\rm 3D}$ survey with no morphological and kinematic
peculiarities and seen at intermediate inclination using the GAlaxy
Surface Photometry 2 Dimensional decomposition algorithm
\citep[GASP2D;][]{Mendez2008, Mendez2014}.
We derived the structural parameters of each galaxy assuming that its
surface brightness distribution was the sum of a S\'ersic bulge
\citep{Sersic1968} and a double-exponential disk \citep{Pohlen2006}, 
as described in \citet{Mendez2017}.
We assumed the isophotes of the bulge and disk to be elliptical,
centered onto the galaxy center, and with constant position angle of
the major axis and constant apparent axial ratio.
GASP2D returns the best-fitting values of the structural parameters of
the bulge (effective surface brightness $I_{\rm e}$, effective radius
$r_{\rm e}$, S\'ersic index $n$, position angle ${\it PA}_{\rm
  bulge}$, apparent axial ratio $q_{\rm bulge}$), disk (central surface
brightness $I_{\rm 0} $, inner scalelength $h$, outer scalelength
$h_{\rm o}$, break radius $r_{\rm break}$, position angle ${\it PA}_{\rm
  disk}$, apparent axial ratio $q_{\rm disk}$) with a $\chi^2$ minimization by
weighting the surface brightness of the image pixels according to the
variance of the total observed photon counts due to the contribution
of both galaxy and sky. The algorithm accounts also for photon noise,
CCD gain and read-out noise, and image PSF and it excludes masked
pixels from the minimization process.

\begin{figure*}
\centering
\includegraphics[width=15cm]{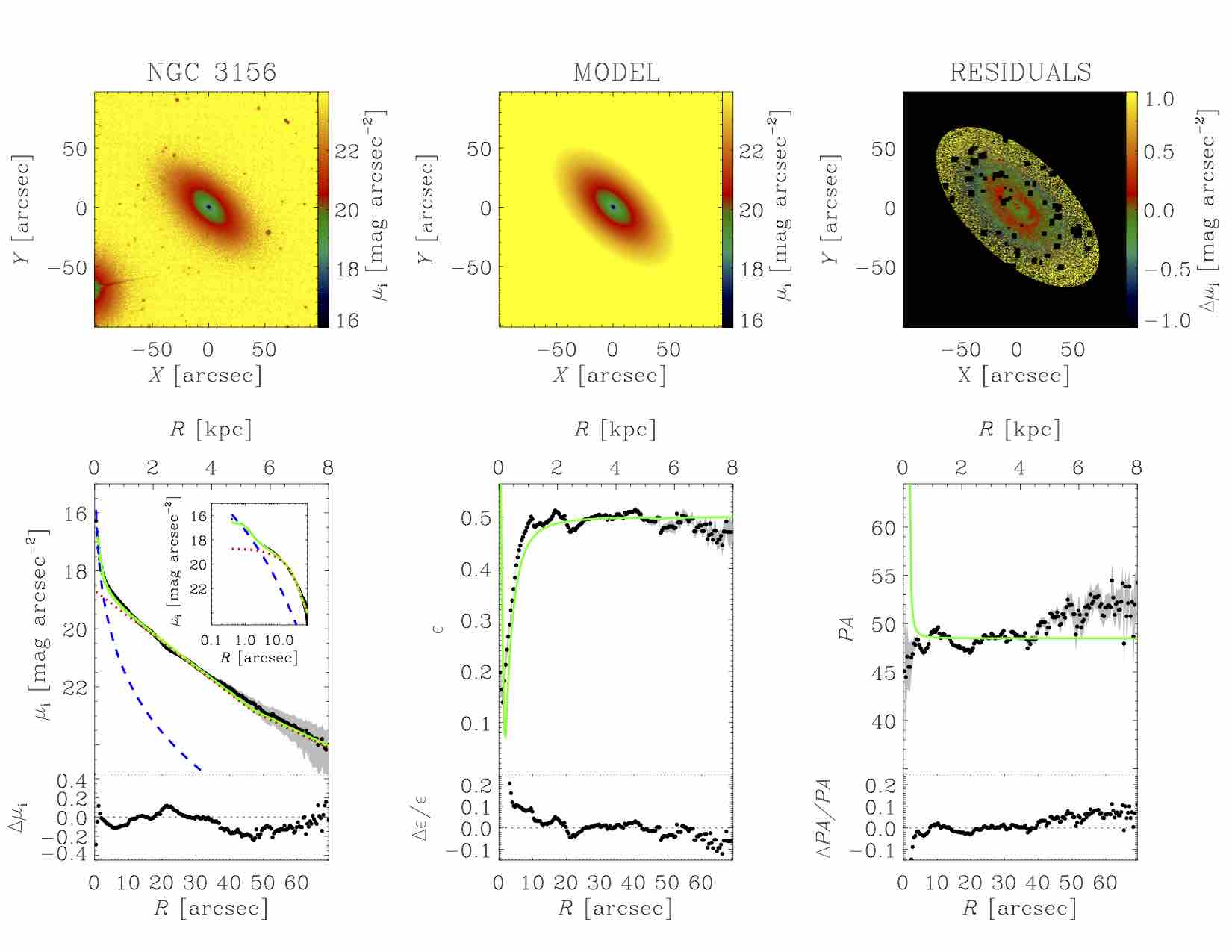}
\caption{Two-dimensional photometric decomposition of the $i$-band
image of the galaxy NGC~3156 as obtained from GASP2D. The upper
panels (\emph{from left to right}) show the map of the observed,
modeled, and residual (observed$-$modeled) surface-brightness
distributions. The field of view is oriented with North up and East
left. The black areas in the residual image correspond to pixels
excluded from the fit. The lower panels (\emph{from left to right})
show the ellipse-averaged radial profile of surface brightness,
position angle, and ellipticity measured in the observed (black dots
with gray error bars) and seeing-convolved modeled image (green
solid line) and their corresponding difference. The
surface-brightness radial profiles of the best-fitting bulge (blue
dashed line) and disk (red dotted line) are also shown in both
linear and logarithmic scale for the distance to the center of the
galaxy.}
\label{fig:decomposition_n3156}
\end{figure*}

We discriminated between elliptical and lenticular galaxies by using
the logical filtering and statistical criteria given in
\citet{Mendez2018} to decide whether adopt or not a disk component to
model the surface brightness distribution of the selected
galaxies. The logical filter tells the difference between elliptical
and lenticular galaxies by choosing the best-fitting model with a physical
meaning. In fact, the lenticular galaxies are supposed to have both a
bulge contributing most of the galaxy surface brightness in the inner
regions and a disk dominating in the galaxy outskirts. The statistical
significance of this analysis relies on the Bayesian information
criterion \citep[BIC;][]{Schwarz1978}. Following the prescriptions of
\citet{Simard2011}, we adopted the $BIC$ parameter
\begin{equation}
BIC = \chi^2  A_{\rm PSF} + k \ln \left(\dfrac{m}{A_{\rm PSF}} \right) \, ,
\end{equation}
where $k$ is the number of free parameters, $m$ is the number of
independent data points, and $A_{\rm PSF}$ is the size area of the
FWHM$_{\rm PSF}$. As in \citet{Mendez2018}, we took advantage of the
simulated mock galaxies to set at $\Delta BIC$ = $BIC$(bulge) $-$
$BIC$(bulge + disk) $> -18$ the threshold that statistically sets the
distinction between elliptical and lenticular galaxies. This led us to
identify three elliptical galaxies, which we rejected.

We scrutinized the residual images obtained after subtracting the
GASP2D model images of the remaining 19 {\em bona fide} lenticulars
from their SDSS images to look for other components than bulge and
disk (i.e., a main bar, large-scale spiral arms). 
Five galaxies showed a weak bar and/or a
faint spiral structure and were discarded. 

Under the assumption of circular and infinitesimally thin disk, 
we calculated the galaxy inclination $\theta = \arccos(q_{\rm disk})$, 
discarding six galaxies because of their high inclination. 

Finally, we visually inspected the optical and near-infrared images of
each of the remaining eight remaining galaxies available in the Hubble
Space Telescope (HST) Science Archive to double check that they did
not host nuclear bars, rings, or spiral arms. 

We report the best-fitting structural parameters and corresponding
errors for the eight sample galaxies, together with their $\Delta BIC$
values in Table~\ref{tab:decomposition}. We derived the errors on the structural
parameters of these galaxies by analyzing the images of a sample of mock
galaxies generated with Monte Carlo simulations and mimicking the
available SDSS images following the procedure described in
\citet{Costantin2017a}. We assumed the mock galaxies to be at a
distance of 27 Mpc, which corresponds to the median distance of our
sample galaxies.

We show in Fig.~\ref{fig:decomposition_n3156} the photometric decomposition
of NGC~3156 as an example and present the results for the other sample
galaxies in Fig.~\ref{fig:appendix}.

\begin{sidewaystable}
\centering
\vspace{-8.5cm}
\caption{Structural parameters of the sample galaxies.}
\begin{adjustbox}{width=\textwidth}
\begin{tabular}{ccccccccccccccccc}
\hline
Galaxy		&	$\mu_e$ 	&	$ \langle\mu_{\rm e}\rangle$    &	$r_{\rm e}$		&	$n$				&	$q_{\rm bulge}$		&	$PA_{\rm bulge}$	&	$\mu_0$ 				&	$h$					&	$h_{\rm out}$		&	$r_{\rm break}$ 		&	$q_{\rm disk}$			&	$PA_{\rm disk}$	& 	$B/T$	& $\Delta BIC$ \\
			&	[mag arcsec$^{-2}$]		& [mag arcsec$^{-2}$]	&	[arcsec]		&					&						&	[$^{\circ}$]			&	[mag arcsec$^{-2}$]		&	[arcsec]				&	[arcsec]			&	[arcsec]				&						&	[$^{\circ}$]			&			\\
(1)			&	(2)					& (3)				& (4)					& (5)					& (6)						& (7)					& (8)						& (9)						& (10)				& (11)					& (12)					& (13) 				& 	(14)		& (15)\\
\hline
NGC~3156		&	20.06 $\pm$ 0.02	& 18.52 $\pm$ 0.05	&	4.7 $\pm$ 0.1		&	5.31 $\pm$ 0.04	&	0.580 $\pm$ 0.002		&	48.7 $\pm$ 0.1		&	18.68 $\pm$ 0.02		&	12.22 $\pm$ 0.07		& 19.6 $\pm$ 0.6 	& 47 $\pm$ 1		&	0.495 $\pm$ 0.002		&	48.5 $\pm$ 0.2		&	0.16		& 103 \\
NGC~3245		&	17.32 $\pm$ 0.01	& 16.42 $\pm$ 0.01	&	4.03 $\pm$ 0.02	&	1.52 $\pm$ 0.01	&	0.730 $\pm$ 0.001		&	175.7 $\pm$ 0.1	&	18.50 $\pm$ 0.01		&	20.61 $\pm$ 0.03		& 34.3 $\pm$ 0.1 	& 78.4 $\pm$ 0.1	&	0.518 $\pm$ 0.001		&	176.5 $\pm$ 0.1	&	0.25		& 49 \\
NGC~3998		&	17.31 $\pm$ 0.01	& 16.20 $\pm$ 0.02	&	5.40 $\pm$ 0.03	&	2.31 $\pm$ 0.01	&	0.838 $\pm$ 0.001		&	131.6 $\pm$ 0.1	&	19.10 $\pm$ 0.01		&	23.76 $\pm$ 0.06		& 40.7 $\pm$ 0.5	& 98.4 $\pm$ 0.8	&	0.790 $\pm$ 0.001		&	139.0 $\pm$ 0.1	&	0.42		& 2 \\
NGC~4578		&	19.32 $\pm$ 0.01	& 18.13 $\pm$ 0.02	&	8.89 $\pm$ 0.06	&	2.71 $\pm$ 0.01	&	0.750 $\pm$ 0.001		&	32.5 $\pm$ 0.1		&	20.37 $\pm$ 0.01		&	32.51 $\pm$ 0.08		&	-			&		-		&	0.686 $\pm$ 0.001		&	30.6 $\pm$ 0.1		&	0.39		& -16 \\
NGC~4690		&	19.37 $\pm$ 0.02	& 17.91 $\pm$ 0.07	&	2.84 $\pm$ 0.09	&	4.52 $\pm$ 0.04	&	0.779 $\pm$ 0.002		&	148.1 $\pm$ 0.1	&	19.37 $\pm$ 0.02		&	10.4 $\pm$ 0.1			& 18.3 $\pm$ 0.5	& 28.6  $\pm$	0.6	&	0.737 $\pm$ 0.002		&	149.2 $\pm$ 0.2	&	0.19		&  30 \\
NGC~5687		&	19.01 $\pm$ 0.01	& 17.92 $\pm$ 0.03	&	6.67 $\pm$ 0.07	&	2.58 $\pm$ 0.01	&	0.686 $\pm$ 0.001		&	103.6 $\pm$ 0.1	&	20.16 $\pm$ 0.01		&	24.26 $\pm$ 0.08		&	-			&		-		&	0.622 $\pm$ 0.001		&	101.2 $\pm$ 0.1	&	0.40		& -17 \\
NGC~6149		&	18.79 $\pm$ 0.02	& 17.65 $\pm$ 0.07	&	3.0 $\pm$ 0.1		&	2.49 $\pm$ 0.03	&	0.691 $\pm$ 0.002		&	21.2 $\pm$ 0.1		&	19.50 $\pm$ 0.02		&	8.85 $\pm$ 0.05		&	-			&		-		&	0.660 $\pm$ 0.002		&	18.9 $\pm$ 0.2		&	0.39		&  0 \\
NGC~7457		&	21.41 $\pm$ 0.01	&  19.92 $\pm$ 0.02 		& 21.8 $\pm$ 0.1	&	4.86 $\pm$ 0.01	&	0.645 $\pm$ 0.002 	&	128.9 $\pm$ 0.2		&	19.38 $\pm$ 0.06		&	28.01 $\pm$ 0.05		&	-			&		-		&	0.511 $\pm$ 0.001		&	125.0 $\pm$ 0.1	&	0.32		& 14 \\
\hline
\end{tabular}
\end{adjustbox}
\begin{minipage}{239mm} 
{\bf Notes.\/} Column (1): galaxy name. Columns (2), (3), (4), (5), (6) and (7): surface
brightness at effective radius, mean surface brightness within
effective radius, effective radius, S\'ersic index, apparent axial
ratio, and position angle of the bulge, respectively. Columns (8), (9), (10)
(11), (12), and (13): central surface brightness, inner scalelength radius, 
outer scalelength radius, break radius, apparent
axial ratio, and position angle of the disk, respectively. Column (14): bulge-to-total luminosity ratio.
Column (15): $\Delta BIC = BIC$(bulge) $-$ $BIC$(bulge + disk).
\end{minipage}
\label{tab:decomposition}
\end{sidewaystable}

\section{Bulge intrinsic shape}
\label{sec:shape}

We constrained the intrinsic shape of our sample bulges with the
statistical method presented in \citet{Mendez2010} and revised in
\citet{Costantin2017b} using our code galaXYZ written in
IDL\footnote{Interactive Data Language is distributed by ITT Visual
  Information Solutions. It is available from
  \url{http://www.ittvis.com}.}.  In summary, we first assumed that
the bulge can be modeled by a triaxial ellipsoid with an equatorial
axial ratio $B/A$ and flattening $C/A$ that shares both the same
equatorial plane and center of the disk.  Secondly, the disk component
is considered to be an oblate ellipsoid with an intrinsic flattening
described by a normal distribution with
$\langle q_{0, \, \rm disk} \rangle = 0.267 \pm 0.102$
\citep{Rodriguez2013}. It is worth noting that the
inclination of our sample galaxies is in the range $25^{\circ} <
\theta < 65^{\circ}$ for which the intrinsic shape of the bulge can be
successfully constrained with galaXYZ \citep{Costantin2017b}.

\begin{figure}
\centering \includegraphics[width=0.49\textwidth, trim=0.2cm 0.2cm 1.5cm
1cm, clip=true]{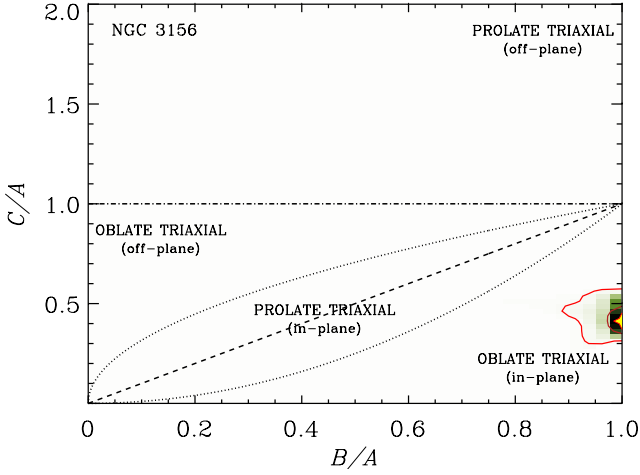}
\caption{Distribution of the intrinsic axial ratios $B/A$ and $C/A$ of
the bulge of NGC~3156.  The yellow star corresponds to the most
probable values of $B/A$ and $C/A$.  The inner and outer red solid
contours respectively encompass the 68.3\% and 95.4\% of the
realizations of ($B/A$, $C/A$) consistent with the geometric
parameters of bulge and disk measured from our photometric
decomposition. Different lines mark the regimes of oblate-triaxial (in-plane),
oblate-triaxial (off-plane), prolate-triaxial
(in-plane), and prolate-triaxial (off-plane).}
\label{fig:shape_n3156}
\end{figure}

We show in Fig.~\ref{fig:shape_n3156} the probability distribution of
$B/A$ and $C/A$ of the bulge in NGC~3156 as an example, while the
remaining sample bulges are presented in Fig.~\ref{fig:appendix}. We
list the most probable values of $B/A$ and $C/A$ of the sample bulges
in Table~\ref{tab:probshape}.

As in \citet{Mendez2018b}, we discriminated bulges according to
their intrinsic shape among
oblate-triaxial and prolate-triaxial ellipsoids,
following the description proposed by \citet{Franx1991}.
Each class contains in-plane systems, which are flattened with respect to the disk equatorial plane,
and off-plane systems, which are elongated along the polar axis.
Special cases of this description include spherical ($A = B = C$), oblate axisymmetric ($B = A$),
and prolate axisymmetric ($B = C$) spheroids.

\begin{table}
\caption{Most probable intrinsic shape of our sample bulges.}
\centering
\begin{tabular}{cccccc}
\hline
Galaxy & $B/A$	&	$C/A$ &  $P({C/A < 0.369})_{\rm bulge}$ 		&  $T_1$  \\
(1)    & (2)                 & (3)      & (4)  & (5) \\
\hline
NGC~3156	&	1.00		&	0.41		&		15\%			&		20\%	\\
NGC~3245	&	1.00		&	0.61		&		0\%			&		 4\%	\\
NGC~3998	&	0.94		&	0.59		&		15\% 		&		25\%	\\
NGC~4578	&	1.00		&	0.46		&		10\%			&		20\%	\\
NGC~4690	&	1.00		&	0.46		&		10\%			&		19\%	\\
NGC~5687	&	0.94		&	0.51		&		3\% 			&		18\%	\\
NGC~6149	&	0.96		&	0.36		&		45\%			& 		51\%  \\
NGC~7457	&	0.91		&	0.49		&		8\%			& 		22\% 	\\
\hline
\end{tabular}
\begin{minipage}{8.5cm} 
{\bf Notes.\/} Column (1): galaxy name. Columns (2), (3): most probable intrinsic axial ratios of the bulge.
Column (4): probability that the galaxy hosts an oblate bulge ($B/A > 0.85$)
with an intrinsic flattening ($C/A$) less than 0.369.
Column (5): $T_1$ test statistics as defined in Sect.~\ref{sec:shape}.
\end{minipage}
\label{tab:probshape}
\end{table}

Considering the intrinsic flattening of disks in nearby
galaxies $\langle q_{0, \, \rm disk} \rangle$, we calculated the
probability of having a bulge as flattened as a disk component. 
Firstly, we derived the probability $P(C/A < 0.369)_{\rm bulge}$  for each sample
bulge to be oblate ($B/A > 0.85$) and have a flattening $C/A<0.369$ 
by taking into account the probability density function (PDF) of its axial
ratios in the ($B/A$, $C/A$) diagram. This allows us to
identify the more flattened bulges in our sample.
Indeed, it is worth noting that this probability does not discriminate bulge types,
but allows to characterize the more flattened systems comparing them with the intrinsic
flattening of nearby disks.
Secondly, we built up a statistical hypothesis test for discerning 
between disk-like bulges (null hypothesis $H_0$) and classical 
bulges (alternative hypothesis $H_1$) based on their $C/A$ distribution.
We considered as disk-like bulges the systems with an
intrinsic flattening $(C/A)_{\rm bulge}$ similar to that of nearby disks.
We set the test statistic $T_1$, which calculates the shared area $A_{\rm disk-like}$
between the PDF of the intrinsic flattening of nearby disks 
and the marginalized PDF over $B/A$ of a sample bulge, where 
the full $A_{\rm disk-like}$ is considered in the region 
$(C/A)_{\rm bulge} < \langle q_{0, \, \rm disk} \rangle$.
To set the statistical limits of the hypothesis test, 
we generated a control sample of mock disk-like bulges,
taking advantage of the PDFs of the bulge intrinsic shape in \citet{Costantin2017b}.
For this purpose, we replicated and marginalized 5000 of those PDFs over $B/A$ and
centered each of them on a random value of $q_{0, \, \rm disk}$ sampled from the normal disk distribution,
i.e., we created a control sample of disk-like bulges
with marginalized PDFs comparable with those measured with our galaXYZ code.
Therefore, applying the test statistics to the control sample of mock disk-like
bulges, we were able to set the limit that corresponds to a statistical test at 90\% confidence level.
Thus, this allowed us to identify classical bulges (i.e., rejecting $H_0$ in favor of $H_1$) as those having
$T_1 < 34\%$ at 90\% confidence level (Table~\ref{tab:probshape}).

The results are given in Table~\ref{tab:probshape}. 
We found that NGC~6149 is the most flattened oblate
bulge of our sample. 
Moreover, since it fails our statistical test,
it is the only candidate to be disk-like according to 
its three-dimensional shape.

\section{Integral field spectroscopy}
\label{sec:spectroscopy}

For each sample galaxy, we computed the values of the bulge diagnostics listed in
Sect.~\ref{sec:criteria} and based on stellar kinematics and
line-strength indices for each galaxy of the sample. To this aim, we
took advantage of the two-dimensional maps of the line-of-sight (LOS)
stellar velocity and velocity dispersion and of the equivalent width
of the \Mg\ and \Fe\ line-strength indices provided by the 
ATLAS$^{\rm 3D}$ survey\footnote{The ATLAS$^{\rm 3D}$ data are 
available at \url{http://www-astro.physics.ox.ac.uk/atlas3d/}.} 
\citep[see][for all details]{Emsellem2004,
Cappellari2011, McDermid2015}. 

\subsection{Stellar kinematics}
\label{subsec:kin}

For each galaxy, we calculated the luminosity-weighted LOS velocity
dispersion within the bulge effective radius $\sigma_{\rm e}$ as
\begin{equation}
\sigma_{\rm e} = \dfrac{\sum_{i=1}^N F_i \sigma_i}{\sum_{i=1}^N F_i}  \, ,
\end{equation}
where $\sigma_i$ is the LOS velocity dispersion and $F_i$ is
flux of $i$-th Voronoi bin within the elliptical aperture with
semi-major axis $r{_{\rm e}}$, axial ratio $q_{\rm bulge}$, and
position angle $PA_{\rm bulge}$.
We defined the central velocity dispersion $\sigma_0$ 
(diagnostics (II-1)$_{\rm C}$) as the velocity
dispersion within an elliptical aperture of radius $r{_{\rm e}/10}$. 
We calculated it from $\sigma_{\rm e}$ as
\begin{equation}
\sigma_0 =  \sigma_{r_{\rm e}/10} = {\sigma_{\rm e}} 10^{0.055\pm0.020} \, ,
\end{equation}
using the aperture correction derived for ETGs by
\citet{FalconBarroso2016}, who analyzed of the stellar kinematics of
the galaxies mapped by the Calar Alto Legacy Integral Field Area
Survey (CALIFA) data release 3 \citep{Sanchez2016}.

Similarly to \citet{Fabricius2012}, we calculated the
luminosity-weighted $\langle v^2\rangle/\langle \sigma^2\rangle$ 
(diagnostics (I-4)$_{\rm C, D}$) as
\begin{equation}
\left. \dfrac{\langle v^2\rangle}{\langle \sigma^2\rangle}
  \right\vert_{25\%} =
  \dfrac{\sum_{i=1}^N F_i v_i^2}{\sum_{i=1}^N F_i \sigma_i^2}  \, ,
\end{equation}
where $F_i$ is the flux of the $i$-th Voronoi bin within the circular
corona with a minimum radius $r_{\rm min} = {FWHM}_{\rm PSF}$ to
minimize the blurring effects of the PSF and a maximum radius
$r_{\rm max} = r_{25\%}$ defined as the radius where the surface brightness
contribution of the bulge exceeds that of the disk by 25\%, that is
\begin{equation}
I_{\rm e} \, e^{b_n} \, e^{- \left(\dfrac{b_n \, r_{25\%}}{r_{\rm e}}\right)^{1/n}}  =
  1.25 \, I_{0, \, \rm disk} \, e^{-\left(\dfrac{r_{25\%}}{h}\right)} \, .
\end{equation}
For future reference, we also provided the luminosity-weighted
$\langle v^2\rangle/\langle \sigma^2\rangle |_{\rm e}$ value inside
the bulge effective radius as
\begin{equation}
\left.\dfrac{\langle v^2\rangle}{\langle \sigma^2\rangle} \right\vert_{{\rm e}} =
  \dfrac{\sum_{i=1}^N F_i v_{{\rm corr}, i}^2}{\sum_{i=1}^N F_i \sigma_i^2}  \, ,
\end{equation}
where $v_{{\rm corr}, i}$ is the inclination-corrected velocity,
$\sigma_i$ is the LOS velocity dispersion, and $F_i$ is the
flux of the $i$-th Voronoi bin in the same elliptical aperture adopted
for measuring $\sigma_{\rm e}$ \citep{Binney2005}.
  
Finally, we derived the logarithmic slope $\gamma$ of the velocity
dispersion radial profile  (diagnostics (I-4)$_{\rm C, D}$) as
\begin{equation}
\gamma = \left. \left\langle \dfrac{{\rm d}\log(\sigma)}{{\rm d}\log(r)}
  \right \rangle \right \vert_{25\%} \, .
\end{equation}
We made sure to avoid a dependence on the particular binning scheme of
each kinematic data set by using a circular radial binning of five
equally sized bins in $\log(r)$ \citep{Fabricius2012}. 

\begin{table*}
\caption{Kinematic parameters and line-strength indices for the sample galaxies.}  
\centering
\begin{adjustbox}{width=\textwidth}
\begin{tabular}{cccccccccccc}
\hline
Galaxy & $r_{25\%}$ & $\sigma_{r_{\rm e}/10}$  & $\sigma_{\rm e}$ & $\langle v^2\rangle/\langle \sigma^2 \rangle$|$_{\rm 25\%}$   & $\langle v^2\rangle/\langle \sigma^2 \rangle$|$_{\rm e}$   & Mg$_b$   &  Mg$_{b, {\rm e}}$ & Fe$_{5015}$  & Fe$_{5015, {\rm e}}$   & $\gamma$ \\
       & [arcsec] & [km s$^{-1}$] & [km s$^{-1}$] &   & & [\AA] & [\AA]     & [\AA] & [\AA]  &        \\
(1)    & (2)                 & (3)      & (4)   & (5)      & (6)    & (7) & (8)      & (9)    & (10) & (11) \\
\hline
NGC~3156	& 2.28	&	71	$\pm$ 9	&	 62	$\pm$ 8	&	0.11	&	0.16	&	1.50	$\pm$ 0.03	&	1.62	$\pm$ 0.06  	&	3.65	$\pm$ 0.06	&	3.4	$\pm$ 0.1 	&	-0.11 \\
NGC~3245	& 7.03	&	229	$\pm$ 5	&	 202	$\pm$ 3	&	0.19	&	0.16	&	4.10	$\pm$ 0.03	&	3.97 $\pm$ 0.03  	&	5.23	$\pm$ 0.06	&	5.01	$\pm$ 0.08	&	-0.18	 \\
NGC~3998	& 12.93	&	282	$\pm$ 6	&	 249	$\pm$ 4	&	0.22	&	0.16	&	4.67	$\pm$ 0.03	&	4.52 $\pm$ 0.05	&	1.39	$\pm$ 0.05	&	3.94	$\pm$ 0.09 	&	-0.14	 \\
NGC~4578	& 16.03	&	111	$\pm$ 7	&	 98	$\pm$ 6 	&	0.32	&	0.33	&	4.36	$\pm$ 0.03	&	3.70 $\pm$ 0.09	&	5.59	$\pm$ 0.06	&	4.7	$\pm$ 0.2 	&	-0.11	\\
NGC~4690	& 2.94	&	127	$\pm$ 9	&	 112	$\pm$ 8 	&	0.02	&	0.02	&	3.31	$\pm$ 0.07	&	3.08 $\pm$ 0.08	&	4.5	$\pm$ 0.1		&	4.3	$\pm$ 0.2		&	-0.04 \\
NGC~5687	& 12.14	&	193	$\pm$ 9	&	 170	$\pm$ 6 	&	0.20	&	0.18	&	4.28	$\pm$ 0.09	&	3.9 $\pm$ 0.1		&	5.4	$\pm$ 0.2		&	4.8	$\pm$ 0.2		&	-0.11	\\
NGC~6149	& 4.68	&	111	$\pm$ 6	&	 98	$\pm$ 5  	&	0.36	&	0.28	&	3.38	$\pm$ 0.04	&	3.24 $\pm$ 0.05	&	4.64	$\pm$ 0.08	&	4.4	$\pm$ 0.1		&	-0.07	\\
NGC~7457 	& 7.62 	& 	70	$\pm$ 11	& 	 62	$\pm$ 10 	&  	0.06	&	0.33	&	2.93	$\pm$ 0.02 	&	2.85 $\pm$ 0.09	&	5.02	$\pm$ 0.04 	&	4.4	$\pm$ 0.2		&	-0.03	\\
\hline
\end{tabular}
\end{adjustbox}
\begin{minipage}{177mm} 
{\bf Notes.\/} Column (1): galaxy name.  
Column (2): radius where the surface brightness contribution of the bulge
exceeds that of the disk by 25\%.
Columns (3), (4): luminosity-weigthed values of LOS velocity dispersion within
an elliptical aperture of semi-major axis $r_{{\rm e}/10}$ and
$r_{\rm e}$, respectively.
Column (5): luminosity-weighted value of $v^2/\sigma^2$ within a circular
corona between $r_{\rm min} = {FWHM}_{\rm PSF}$ and $r_{\rm max} = r_{25\%}$.
Column (6): luminosity-weighted value of $v^2/\sigma^2$ within an elliptical
aperture of semi-major axis $r_{\rm e}$.
Columns (7), (9): luminosity-weigthed values of the \Mg\ and \Fe\ line-strength
indices within a circular aperture of 1.5 arcsec.  
Columns (8), (10): luminosity-weigthed values of the \Mg\ and \Fe\ line-strength
indices within an elliptical aperture of semi-major axis $r_{\rm e}$.
Column (11: logarithmic slope of the radial profile of the LOS velocity
dispersion between $r_{\rm min} = {FWHM}_{\rm PSF}$ and $r_{\rm max} = r_{25\%}$.
\end{minipage}
\label{tab:kin_spec}
\end{table*}

\begin{figure}
\centering \includegraphics[width=0.49\textwidth, trim=0.5cm 2cm 1cm 2.7cm, clip=true]{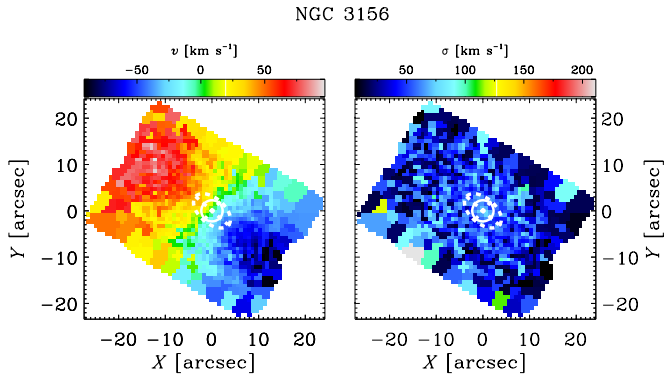}
\caption{Two-dimensional map of the LOS velocity (\emph{left panel})
and velocity dispersion (\emph{right panel}) of the stellar
component of NGC~3156. The white dashed line corresponds to the
elliptical aperture with semi-major axis $r{_{\rm e}}$, axial ratio
$q_{\rm bulge}$, and position angle $PA_{\rm bulge}$. The white
solid line marks the circle with radius $r_{25\%}$. North is up and
east is left.}
\label{fig:kinematics_n3156}
\end{figure}

\begin{figure}
\centering
\includegraphics[width=0.49\textwidth, trim=0.5cm 2.4cm 1cm 2.7cm, clip=true]{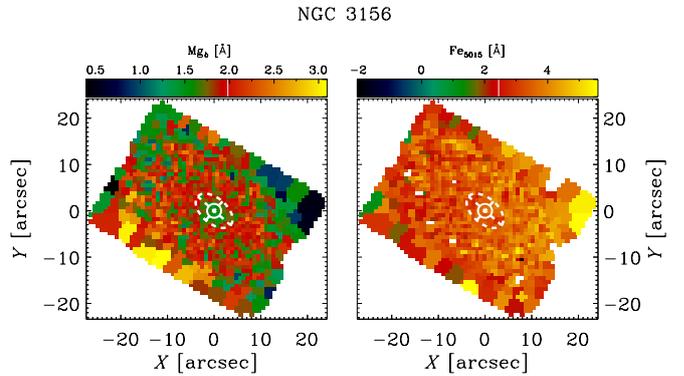}
\caption{Two-dimensional map of the equivalent width of the
  \Mg\ (\emph{left panel}) and \Fe\ (\emph{right panel}) line-strength
  indices of NGC~3156. The white dashed line corresponds to the
  elliptical aperture with semi-major axis $r{_{\rm e}}$, axial ratio
  $q_{\rm bulge}$, and position angle $PA_{\rm bulge}$. The white
  solid line marks the circle with radius 1.5 arcsec. North is up and
 east is left.}
\label{fig:indices_n3156}
\end{figure}

We provide the values of $r_{25\%}$, $\sigma_{r_{\rm e}/10}$,
$\sigma_{\rm e}$, $\langle v^2\rangle/\langle \sigma^2\rangle|_{\rm
  25\%}$, $\langle v^2\rangle/\langle \sigma^2\rangle|_{\rm e}$, and
$\gamma$ in Table~\ref{tab:kin_spec}.  We show in
Fig.~\ref{fig:kinematics_n3156} the stellar kinematics of NGC~3156 as
example and show the remaining galaxies in Fig.~\ref{fig:appendix}.

It is worth noting that the field of view of the stellar kinematic
maps typically encompasses one galaxy effective radius
(Table~\ref{tab:sample}) ensuring the full coverage of the
bulge-dominated region. To give an idea of the bulge size, we overplot
to the stellar kinematic maps the ellipse with semi-major axis
$r{_{\rm e}}$, axial ratio $q_{\rm bulge}$, and position angle
$PA_{\rm bulge}$ within which we calculated $\sigma_{\rm e}$ and
$\langle v^2\rangle/\langle \sigma^2\rangle |_{{\rm e}}$ as well as
the circle with a radius of $r_{25\%}$.  It results that if a galaxy
shows a centrally-peaked velocity dispersion, the increase of the
velocity dispersion is generally confined within the effective radius
of the bulge (e.g., NGC~3245 and NGC~3998).

\subsection{Line-strength indices}

\begin{figure}
\centering \includegraphics[width=0.49\textwidth, trim=0.5cm 0.1cm 1cm
  2cm, clip=true]{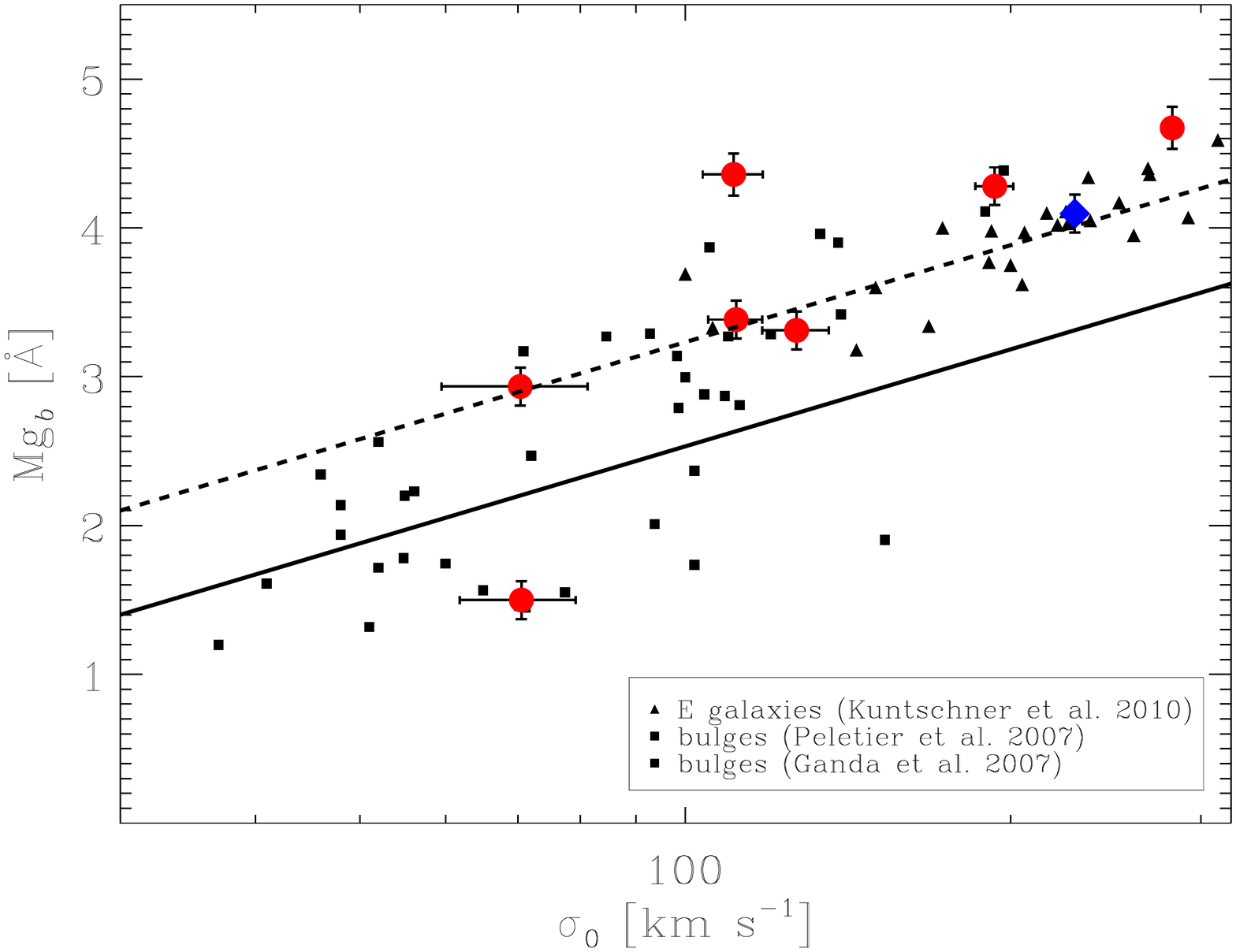}
\caption{\Mg--$\sigma_0$ relation for our sample bulges (large
  circles and diamonds) and the comparison sample of elliptical galaxies (small
  triangles) from \citet{Kuntschner2010} and bulges (small squares)
  from \citet{Peletier2007} and \citet{Ganda2007}. Red and blue
  symbols mark the sample bulges with $n > 2$ and $n < 2$,
  respectively. The black dashed line is the best-fitting relation
  inferred by \citet{Fisher2016} for the elliptical galaxies, while
  the black solid line corresponds to a deviation of 0.7 \AA\ in
  \Mg\ from the best-fitting relation.}
\label{fig:relation1}
\end{figure}

For each galaxy, we derived the luminosity-weighted central values of the
line-strength indices \Mg\ and \Fe\ within a circular aperture of
$1.5$-arcsec radius as 
\begin{equation}
\emph{Index} = \dfrac{\sum_{i=1}^N F_i \emph{Index}_i}{\sum_{i=1}^N F_i}  \, ,
\end{equation}
where $Index_i$ is the index equivalent width and $F_i$ is flux of
the $i$-th Voronoi bin within the selected aperture. This allows to
apply of the bulge diagnostics (I-3)$_{\rm D}$ and (II-3)$_{\rm D}$
related to the properties of their stellar populations.

\citet{Peletier2007} measured the line-strength indices within a
circular aperture of 1.2-arcsec radius for 24 bulges of early-type
spiral galaxies, while \citet{Ganda2007} measured them within a
circular aperture of 1.5-arcsec radius for 18 bulges of late-type
spiral galaxies. \citet{Fisher2016} combined these measurements,
stressing that no classical bulge has \Mg$ < 2.35$ \AA\ or \Fe$ <
3.97$ \AA. It has to be strongly remarked that these thresholds were
chosen once the bulges were already \emph{a priori} classified as
classical or disk-like by analyzing their visual morphological
classification (diagnostics (I-1)$_{\rm C, D}$) and/or S\'ersic index
(diagnostics (I-2)$_{\rm C, D}$). For future reference, we also
calculated the luminosity-weighted values \Mge\ and \Fee\ of the
line-strength indices \Mg\ and \Fe\ in the same elliptical aperture
adopted for measuring $\sigma_{\rm e}$.

We listed the measured values of \Mg , \Mge , \Fe , and \Fee\ of the
sample galaxies in Table~\ref{tab:kin_spec}. We plot in
Fig.~\ref{fig:indices_n3156} the two-dimensional map of the equivalent
width of the \Mg\ and \Fe\ line-strength indices of NGC~3156 as an
example, while the remaining galaxies are shown in
Fig.~\ref{fig:appendix}. We overplot the ellipse within which we
calculated $\sigma_{\rm e}$, as well as the circle with a radius of
1.5 arcsec. We found that the bulges of NGC~3156 and NGC~3998 have
\Fe$ < 3.95$ \AA, while only the bulge of NGC~3156 also presents \Mg$
< 2.35$ \AA.

\begin{figure}
\centering \includegraphics[width=0.49\textwidth, trim=0.5cm 0.1cm 1cm
  2cm, clip=true]{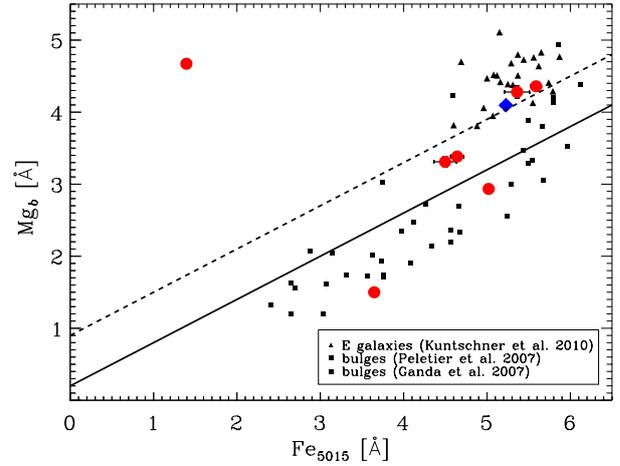}
\caption{As in Fig.~\ref{fig:relation1}, but for the
  \Mg--\Fe\ relation.}
\label{fig:relation2}
\end{figure}

Furthermore, we considered the \Mg--$\sigma_0$ and
\Mg--\Fe\ relationships  (diagnostics (I-3)$_{\rm D}$) 
following \citet{Fisher2016}. We adopted as
comparison sample the elliptical galaxies studied by \citet[][where
  $\sigma_0 = \sigma_{r_{\rm e}/8}$]{Kuntschner2010} and the bulges
from \citet[][where $\sigma_0 = \sigma_{1.2 \, \rm
    arcsec}$]{Peletier2007}, and \citet[][where $\sigma_0 =
  \sigma_{1.5 \, \rm arcsec}$]{Ganda2007}. We plot the two
relationships with the best fit to the elliptical galaxies by
\citet{Kuntschner2010} inferred from \citet{Fisher2016} in
Figs.~\ref{fig:relation1} and \ref{fig:relation2}, respectively.

We distinguished our sample bulges according to their S\'ersic index
between $n > 2$ and $n < 2$ as proposed by \citet{Fisher2016} and we
found that only the bulge of NGC~3156 is below the line that is supposed to
separate the two bulge classes in both diagrams, 
with the bulge of NGC~7457 also meeting the disk-like requirement for the \Mg--\Fe\ relation.
The bulge of NGC~3998
is characterized by a very small value of \Fe.  However, considering
\Fee\  it moves towards the \Mg--\Fe\ relation.


\section{Scaling relations}
\label{sec:scaling}

Following \citet{Costantin2017a}, we built the FPR
\citep{Djorgovski1987, Cappellari2013}, 
FJR \citep{Faber1976, FalconBarroso2011}, and 
KR \citep{Kormendy1977, NigocheNetro2008}
using the photometric ($r_{\rm e}$, $\langle \mu_{\rm e} \rangle$, and
$M_{{\rm bulge}, i}$) and kinematic ($\sigma_{\rm e}$) parameters of
elliptical galaxies and bulges from \citet{Gadotti2009} and
\citet{Oh2011}, respectively.  

\begin{figure}
\centering \includegraphics[width=0.49\textwidth, trim=0cm 0.2cm 1.8cm
  2.2cm, clip=true]{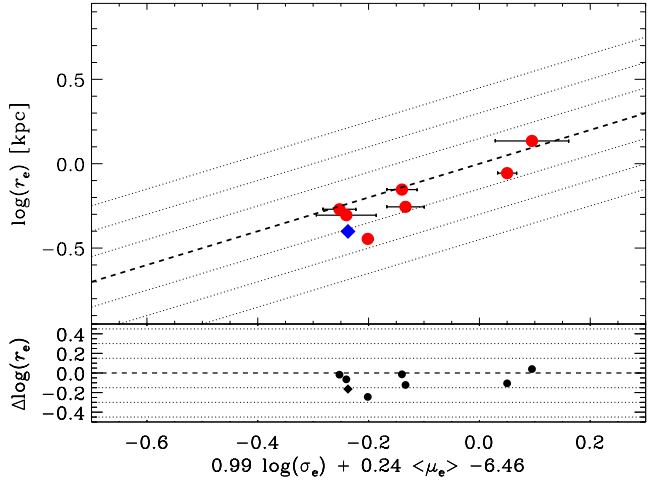}
\caption{Fundamental plane relation for our sample bulges. Red and
  blue symbols mark the sample bulges with $n > 2$ and $n < 2$,
  respectively. The black dashed line corresponds to the best-fitting
  relation derived by \citet{Costantin2017a}. The black dotted lines
  show the $1$ rms, $2$ rms, and $3$ rms deviation in $\log{(r_{\rm
      e})}$ regions, respectively.}
\label{fig:fundamental_plane}
\end{figure}

\begin{figure}
\centering \includegraphics[width=0.49\textwidth, trim=0cm 0cm 1.8cm
  2.2cm, clip=true]{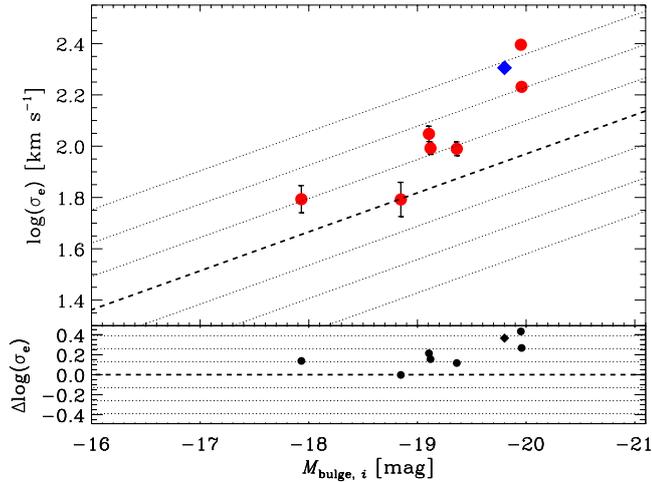}
\caption{As in Fig.~\ref{fig:fundamental_plane}, but for the Faber-Jackson relation and
  with the rms deviation in $\log{(\sigma_{\rm e})}$ from the best
  fitting-relation.}
\label{fig:faber_jackson}
\end{figure}

\begin{figure}
\centering \includegraphics[width=0.49\textwidth, trim=0cm 0cm 1.8cm
  2.2cm, clip=true]{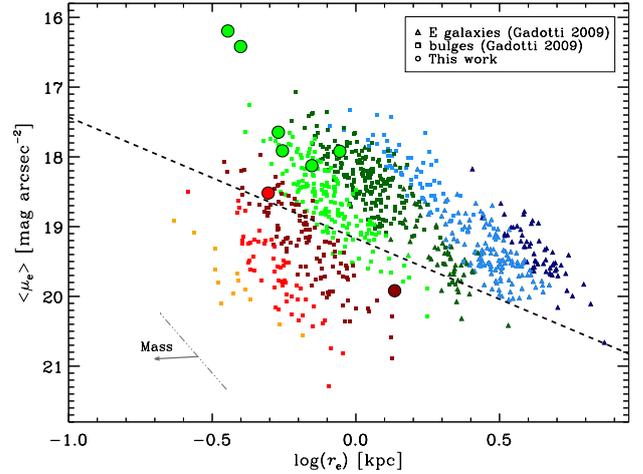}
\caption{Kormendy relation for the bulges of our (large circles) and
comparison sample from \citet[][small triangles: elliptical
galaxies, small squares: bulges]{Gadotti2009}. The bulges are
divided according to their absolute magnitude in the following bins:
$M_i< -22$ mag (dark blue), $M_i = [-22, -21]$ mag (light blue),
$M_i = [-21, -20]$ mag (dark green), $M_i = [-20, -19]$ mag (light
green), $M_i = [-19, -18]$ mag (dark red), $M_i = [-18, -17]$ mag
(light red), and $M_i = [-17, -16]$ mag (orange).  The black dashed
line separates classical from disk-like bulges according to
\citet{Gadotti2009}.  The black dash-dotted line gives the slope of
the relation for the magnitude bin $M_i = [-20, -19]$ mag, while the arrow
indicates the trend for decreasing masses
\citep{NigocheNetro2008}.}
\label{fig:kormendy}
\end{figure}

We combined the photometric and kinematic properties of our bulges to
study their location in the FPR (Fig.~\ref{fig:fundamental_plane};
diagnostics (III-2)$_{\rm C}$) and FJR (Fig.~\ref{fig:faber_jackson};
diagnostics (II-4)$_{\rm D}$), using the best-fitting relations
provided by \citet{Costantin2017a}
\begin{equation}
\log(r_{\rm e}) = 0.99 \log(\sigma_{\rm e}) + 0.24  \langle \mu_{\rm e} \rangle - \, 6.46
\end{equation}
and
\begin{equation}
\log(\sigma_{\rm e}) = -0.152 (\pm 0.003) M_{i} - 1.07 (\pm 0.07) \, ,
\end{equation}
respectively. We found that none of our sample bulges is a
low-$\sigma$ outlier to either the FPR or the FJR. However, we noticed
that our bulges are located systematically below the FPR best-fitting
line and systematically above the FJR best-fitting line, even if they
are consistent with their global trends within the errors. Only the
bulge of NGC~3998 deviates more then $3\sigma$ in $\log\sigma_{\rm e}$
from the FJR.  We investigated the position of our sample bulges in
the KR (Fig.~\ref{fig:kormendy}; diagnostics (II-1)$_{\rm C}$), taking
advantage of the equation provided by \citet{Gadotti2009}
\begin{equation}
\langle \mu_{\rm e} \rangle = 1.74 \, \log(r_{\rm e}) + 19.17 \, ,
\end{equation}
to separate classical from disk-like bulges. 
We found that all our bulges are consistent with the magnitude trend
highlighted by \citet{NigocheNetro2008} and \citet{Costantin2017a},
discriminating between less and more massive bulges. As a consequence,
using the KR to separate bulge types results in classifying less
massive bulges as disk like. Therefore, in the low-mass regime even
the most luminous bulges are supposed to be characterized by disk-like
properties. In addition, we noticed that only the bulge of NGC~7457 is below
the boundary line of the disk-like systems.

\section{Discussion}
\label{sec:discussion}

The observed properties of nearby bulges somehow
preserve the relic of their formation and evolution. Thus, different
observed properties are expected to be associated to various formation
scenarios and, by consequence, to different bulge types. As proposed
by \citet{Fisher2016} and \citet{Kormendy2016}, the proper strategy would be to compare different
diagnostics to take advantage of many observed properties. This is not
usually done in the literature, since it is much easier to consider only
one (or few) diagnostics, leading to misunderstandings due 
to misinterpretations of observational data and numerical results.
For example, some recent numerical simulations have apparently challenged
the current understanding of the relationships between the
classification and formation of bulges. By fine-tuning simulation
parameters, it has been possible to build classical bulges by disk
instabilities and disk-like bulges from minor mergers \citep{Keselman2012, 
Brooks2016, Weinzirl2009}. 
However, the classical/disk-like bulge classification of these 
simulated galaxies is done using only the bulge S\'ersic index as diagnostics. 
Thus, if different formation processes lead to the same
observed property (as it results for the bulge S\'ersic index), this
makes the adopted diagnostics not fully reliable for discriminating the 
formation scenario of bulges and therefore not suitable for their
classification.

{In this context, the question is whether any, or any combination, of the
diagnostics proposed by \citep{Fisher2016} could mark a
distinction between the formation processes of bulges, or whether
other new diagnostics could provide such a distinction.
Thus, we} compared the photometric, kinematic, and stellar population
properties we derived for the sample bulges with the observational
criteria given in Sect.~\ref{sec:criteria} to identify classical
(Table~\ref{tab:classical}) and disk-like bulges
(Table~\ref{tab:disk}). The proposed diagnostics are potentially 
good in a statistical sense, while they result 
uncertain for individual galaxies.

\begin{table*}
\caption{Classification criteria of classical bulges.}
\centering
\begin{tabular}{ccccccccc|c}
\hline
Galaxy & Morph.               & $n$       & \multicolumn{2}{c}{Line-str.~rel.}    &  $\gamma$    & $\sigma_{0}$ & FPR   & Classical bulge & 3D shape \\
  & (I--1)$_{\rm C}$                & (I--2)$_{\rm C}$       &  \multicolumn{2}{c}{(I--3)$_{\rm C}$}   & (I--4)$_{\rm C}$    & (II--1)$_{\rm C}$ & (III--1)$_{\rm C}$ &   &   \\
(1)    & (2)                 & (3)      & \multicolumn{2}{c}{(4)}   & (5)      & (6)    & (7) & (8) & (9)\\
\hline
NGC~3156	& yes	&	yes	&	no	& no		&	yes	&	no	&	yes 		&	 ?		&	 yes	\\
NGC~3245	& yes	&	no	&	yes	& yes	&	yes	&	yes	&	yes 		&	 yes		&	 yes	\\
NGC~3998	& yes	&	yes	&	yes	& yes	&	yes	&	yes	&	yes 		&	 yes		&	 yes	\\
NGC~4578	& yes	&	yes	&	yes	& yes 	&	yes	&	no	&	yes 		&	 yes		&	 yes	\\
NGC~4690	& yes	&	yes	&	yes	& yes	&	no	&	no	&	yes 		&	 ?		&	 yes	\\
NGC~5687	& ...		&	yes	&	yes	& yes	&	yes	&	yes	&	yes 		&	 yes		&	 yes	\\
NGC~6149	& ...		&	yes	&	yes	& yes	&	no	&	no	&	yes 		&	 ?		&	 no	\\
NGC~7457	& yes		&	yes	&	yes 	& no		&	no	&	no	&	yes 		&	 ?		&	 yes	\\
\hline
\end{tabular}
\begin{minipage}{180mm} 
{\bf Notes.\/} Column (1): galaxy name.
Column (2): morphological features from HST images (... = unavailable HST image). Column (3): S\'ersic index $n > 2$.
Column (4): consistency with \Mg--\Fe and \Mg--$\sigma_0$ correlations, respectively.
Column (5): velocity dispersion gradient $\gamma < -0.1$.
Column (6): central velocity dispersion $\sigma_0 > 130$ km s$^{-1}$.
Column (7): consistency with FPR. Column (8): bulge classification according to the
observational diagnostics listed in Cols. (2)-(7) and explained in Sect.~\ref{sec:criteria} (? = uncertain).
Column (9): thick oblate spheroid: either $B/A > 0.85$ \& $C/A > 0.37$ or $B/A < 0.85$ \& any $C/A$.
\end{minipage}
\label{tab:classical}
\end{table*}

\begin{table*}
\caption{Classification criteria of disk-like bulges.}
\centering
\begin{tabular}{cccccccccccc|c}
\hline
Galaxy & Morph.                & $n$       &  \multicolumn{2}{c}{Line-str.~rel.}   & \multicolumn{2}{c}{Kinematics}    & KR & \multicolumn{2}{c}{Line-str.~ind.} & FJR & Disk-like bulge & 3D shape  \\
 & (I--1)$_{\rm D}$                & (I--2)$_{\rm D}$       &  \multicolumn{2}{c}{(I--3)$_{\rm D}$}   & \multicolumn{2}{c}{(I--4)$_{\rm D}$}    & (II--1)$_{\rm D}$ & \multicolumn{2}{c}{(II--3)$_{\rm D}$} & (II--4)$_{\rm D}$ &  &   \\
(1)    & (2)                 & (3)      & \multicolumn{2}{c}{(4)}   & (5)      & (6)    & (7) & (8) & (9) & (10) & (11)  & (12)\\
\hline
NGC~3156	 	& no 			&	no		&	yes	&	yes	&	no		&	no		&	no		&	yes		&	yes		&	no	& ?	&	no	 		 \\
NGC~3245	 	& no			&	yes		&	no	&	no	&	no		& 	no		&	no		&	no		&	no		&	no	& no	&	no 			 \\
NGC~3998	 	& no			&	no		&	no	&	no	&	no		&	no		&	no		&	no		&	yes		&	no	& no	&	no	 		 \\
NGC~4578		& no			& 	no 		&	no	&	no	&	no		&	no		&	no		&	no		&	no		&	no	& no	&	no 			 \\
NGC~4690		& no			&	no		&	no	&	no	&	no		&	yes		&	no		&	no		&	no		&	no	& ?	&	no 			 \\
NGC~5687		& ...			&	no		&	no	&	no	&	no		&	no		&	no		&	no		&	no		&	no	& no	&	no 			 \\
NGC~6149		& ...			&	no		&	no	&	no	&	yes		&	yes		&	no		&	no		&	no		&	no	& ?	&	yes			 \\
NGC~7457		& no 	&	no		&	yes	&	no	&	no		&	yes		&	yes 		&	 no		&	no 		&	no 	& ?	& 	no			 \\
\hline
\end{tabular}
\begin{minipage}{180mm} 
{\bf Notes.\/} Column (1): galaxy name.
Column (2): morphological features from HST images (... = unavailable HST image).
Column (3): S\'ersic index $n < 2$.
Column (4): deviation from \Mg--\Fe and \Mg--$\sigma_0$ correlations, respectively.
Columns (5), (6): $\langle v^2 \rangle/\langle\sigma^2 \rangle|_{25\%} \ge 0.35$ and
velocity dispersion gradient $\gamma \ge -0.1$, respectively.
Column (7): low surface brightness outlier from KR.
Columns (8), (9): line-strength indices (\Mg$ < 2.35$ \AA\, \Fe$ < 3.95$ \AA).
Column (10): low-$\sigma$ outliers from FJR.
Column (11): bulge classification according to the
observational diagnostics listed in Cols. (2)-(7) and explained in
Sect.~\ref{sec:criteria} (? = uncertain).
Column (12): flattened oblate spheroid: $B/A > 0.85$ \& $C/A < 0.37$.
\end{minipage}
\label{tab:disk}
\end{table*}

Only the bulge of NGC~5687 can be unambiguously classified as
classical, since it satisfies all the corresponding criteria and
presents none of the characteristics for being disk-like. The
remaining bulges show a more complex and therefore disputable
behaviour. Nevertheless, we proposed to classify as classical also the bulges of
NGC~3245, NGC~3998, and NGC~4578. Each of them actually misses only
one of the criteria for being classical: the bulge of NGC~3254 has a
small S\'ersic index (diagnostics (I-2)$_{\rm C}$), NGC~3998 does not
follow the same correlations between line-strength indices as
elliptical galaxies (diagnostics (I-3)$_{\rm C}$), and NGC~4578 has a
low central velocity dispersion (diagnostics (II-1)$_{\rm
  D}$). However, performing a bulge classification using only one of
these three diagnostics is quite controversial, because none of them
is effective by itself to mark a clearcut separation between classical
and disk-like bulges.

Firstly, the S\'ersic index is the most extensively adopted
diagnostics to classify bulges \citep[][]{Fisher2008, Neumann2017,
  Kruk2018}, since the bimodal distribution of $n$ is supposed to
separate bulges in classical ($n > 2$) or disk-like ($n < 2$).
However, a physical explanation for this bimodal distribution and for
the empirical boundary line $n = 2$ is not well understood yet
\citep{Fisher2016}. Moreover, many authors pointed out that the
S\'ersic index is prone to misclassifications
\citep[e.g.,][]{Graham2008, Mendez2018}. It is known that mergers can
build bulges with $n < 2$ \citep{ElicheMoral2011, Querejeta2015} and
low-luminosity elliptical galaxies have $n < 2$ or even $n \sim 1$
\citep{Davies1988, Young1994}. Thus, considering as disk-like all the
systems with low S\'ersic index leads to a heterogeneous collection
of bulges with different formation scenarios, rather than singling out
only the bulges built up from disk material during long-lasting
processes.

Secondly, the \Mg\ and \Fe\ line-strength indices and their interplay
in the \Mg--\Fe\ relation are supposed to provide a constraint for the
properties of the stellar population of bulges.  However, their
interpretation leads to contradictory outcomes, which are mostly
inconsistent with those obtained from the analysis of the other
photometric or kinematic parameters. We ascribed this to the variety
of techniques adopted to analyze data and measure the line-strength
indices of the comparison sample.  Indeed, \citet{Fisher2016} combined
information from both \citet{Peletier2007} and \citet{Ganda2007}, even
if they measured the equivalent width of the line-strength indices
within different circular apertures of radius 1.2 and 1.5 arcsec,
respectively.  This does not guarantee a fair comparison of different
bulges, since their physical size is not appropriately taken into
account. Furthermore, the separation of classical and disk-like bulges
in the \Mg--$\sigma_0$ and \Mg--\Fe\ relations (diagnostics
(I-3)$_{\rm D}$) was completely based on empirical results, once
classical and disk-like bulges were already identified according to
their visual morphological classification (diagnostics (I-1)$_{\rm C,D}$) 
and/or S\'ersic index (diagnostics (I-2)$_{\rm C,D}$).  
We found that NGC~3156 hosts the only bulge in our sample
falling in the disk-like region defined from the \Mg--$\sigma_0$ and
\Mg--\Fe\ relations, with the bulge of NGC~7457 also meeting the
disk-like requirement for the \Mg--\Fe\ relation. Nevertheless, both bulges
have $n > 2$, whereas the bulge of NGC~3245 has $n < 2$ and it is
consistent with the expected trends of classical bulges.

\begin{figure}
\centering \includegraphics[width=0.49\textwidth, trim=0.5cm 0.5cm 0.5cm 0.5cm, clip=true]{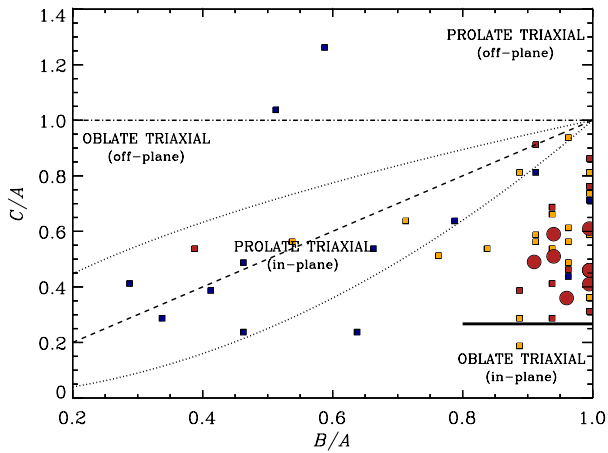}
\caption{As in Fig.~\ref{fig:shape_n3156}, but for the sample bulges
  (large circles) and the bulges of unbarred galaxies studied by
  \citet[][small squares]{Costantin2017b}. Red, yellow and blue
  symbols mark galaxies classified as S0, Sa--Sb, and Sc--Sdm,
  respectively. The thick solid black line corresponds to $\langle q_{0, \, \rm disk} \rangle$
  in \citet{Rodriguez2013}.
  The bulges of NGC~4578 and NGC~4690
  share the same position in the ($B/A$, $C/A$) diagram.}
\label{fig:atlas3d_vs_califa}
\end{figure}

Finally, the stellar kinematics might be a good gauge of the nature of
disk-like bulges, since they are supposed to preserve the properties
of the disks from which they formed. However, we refrained from
considering as disk-like all the bulges with a low velocity
dispersion. Indeed, the velocity dispersion of the bulge only
partially characterizes its dynamical status being just a good proxy
for the mass in dispersion-dominated systems. The fact that only three
out of eight bulges in our sample (NGC~3245, NGC~3998, and NGC~5687)
show a central velocity dispersion $\sigma_{r_{\rm e}/10} > 130$ km
s$^{-1}$ does not rule out the possibility of label other sample
bulges as classical. \citet{Costantin2017a} investigated a sample of
small bulges ($\sigma_{\rm e}\simeq50$ km s$^{-1}$) of late-type
spirals and found they follow the same scaling relations of
ellipticals, massive bulges, and compact early-type galaxies so they
cannot be classified as disk-like systems.

It is worth noting that the failure of the photometric ($n$) and
kinematic ($\sigma$, $\langle v^2\rangle /\langle \sigma^2\rangle$,
and $\gamma$) diagnostics in giving the same classification for our
sample bulges is not surprising. Controversial results have been
obtained when these diagnostics were combined to classify bulges for
which accurate photometric and kinematic measurements were
available. \citet{Mendez2018} found no statistically significant
correlation between the $v/\sigma$ and $n$ regardless of projection
effects by analyzing a sample of lenticular
galaxies. \citet{FalconBarroso2003} studied early-type disk galaxies
and argued that $n$ is not related to $\gamma$. Recently,
\citet{Tabor2017} have performed a spectro-photometric bulge-disk
decomposition of three lenticular galaxies, showing that their bulges are
pressure-supported systems despite they have $n \sim 1$ and some
degree of rotation. On the contrary, \citet{Fabricius2012} claimed
that bulges with $n<2$ of galaxies ranging from lenticular to
late-type spiral galaxies are characterized by an increased rotational
support. The differences probably arise from the fact that bulges of
late-type spiral galaxies are more consistently rotation dominated and have a
lower velocity dispersion than bulges of lenticular galaxies, whose formation
process is more complex \citep{Bekki1998, Governato2009, Mendez2018}.

The position of our sample bulges in the FPR, FJR, and KR confirmed the
recent findings of \citet{Costantin2017a} on scaling relations for
elliptical galaxies and bulges. They claimed that there is a single
population of galaxy spheroids that follow the same FPR
(Fig.~\ref{fig:fundamental_plane}) and FJR
(Fig.~\ref{fig:faber_jackson}) and argued that the mass is responsible
for the smooth transition in the photometric and kinematic properties
from less to more massive bulges.

Photometric, kinematic, and line-strength diagnostics contradict each
others for the bulges of NGC~3156, NGC~4690, NGC~6149, and NGC~7547
making them difficult to be classified. For this reason, we propose
to add a piece of information by considering the bulge three-dimensional shape
as a possible proxy to distinguish their nature. 
Our analysis suggests that all sample bulges are oblate spheroids, 
but only the bulge of NGC~6149 cannot be consider as classical at 90\% C.L.
(Table~\ref{tab:probshape}).
Therefore, we conclude that bulges of NGC~3156, NGC~4690, and NGC~7547 
are most likely classical rather than
disk-like, while NGC~6149 is consider to host a possible disk-like bulge.
As shown in Sect.~\ref{sec:shape}, we claim that our ability to constrain 
the bulge intrinsic shape is not limited to identify only classical bulges.
For comparison, we found that a few oblate bulges from the sample
of unbarred galaxies ranging from S0 to Sdm and taken from the CALIFA survey 
in \citet{Costantin2017b} cannot be classified as classical since they fail 
our statistical test analysis
(Fig.~\ref{fig:atlas3d_vs_califa}).


\section{Conclusions}
\label{sec:conclusions}

Analyzing the SDSS \citep{Ahn2012} and ATLAS$^{\rm 3D}$ 
\citep{Cappellari2011, McDermid2015} datasets, we derived the
photometric and spectroscopic properties of a sample of \emph{bona
  fide} unbarred lenticular galaxies in order to understand whether
they host a classical or a disk-like bulge applying the observational
diagnostics proposed by \citet{Fisher2016}.

We obtained the photometric diagnostics ($n$) of the sample bulges by
performing a two-dimensional photometric decomposition of the SDSS
$i$-band images. We derived the kinematic ($\sigma$, $\langle v^2
\rangle / \langle \sigma^2 \rangle$, and $\gamma$) and line-strength
(\Mg\ and \Fe) diagnostics within different apertures. In addition,
we combined the line-strength indices and velocity dispersion in the
\Mg--\Fe\ and \Mg--$\sigma_0$ relations. Finally, we used the
photometric and kinematic parameters to investigate the location of
the sample bulges in FPR, FJR, and KR built for elliptical and bulges.

We noticed that only sometimes the proposed diagnostics are successful
in identifying classical bulges (Tables~\ref{tab:classical} and
\ref{tab:disk}). As a matter of fact, the kinematic and line-strength
diagnostics provided no clear identification for half of the sample
bulges. This remains true also when we compared the classification
based on the photometric and line-strength diagnostics. 
We derived the intrinsic shape of the sample bulges. All
of them turned out to be thick oblate spheroids, but only NGC~6149
could be considered to most likely host a disk-like bulge. We concluded that all the other
bulges could be classify as classical.
We pointed out that the intrinsic shape of bulges, which reflects the violent/secular
evolution of galaxies, could add a piece of information in
characterizing the bulge types, unveiling extreme cases and tracing a
continuity among them. The analysis of the bulge shape can not replace
a full investigation of all the observed properties. But, it
can be adopted to guess the shape of the gravitational potential in
the center of nearby galaxies without the demand of a full orbital
analysis, which is still not clear whether it would be able to
completely solve this problem \citep{Zhu2018a, Zhu2018b}.

Despite the low number statistics, but taking advantage of the careful
selection of our sample, we concluded that the common practice of
applying the observational diagnostics by \citet{Fisher2016} for
distinguishing bulge types (based on an \emph{a priori} classification
according to their morphology and/or S\'ersic index) has to be
carefully reconsidered. We remarked that, even if each diagnostics
looks well motivated in terms of distinct formation paths of bulges,
their calibration and interplay might result in controversial findings. 
This is a pilot project, which requires further analysis with MUSE spectroscopy 
and HST imaging to improve the data spatial resolution \citep[e.g.,][]{Gadotti2015}, 
and a larger sample that includes barred galaxies and spirals to fine-tune the diagnostics.
We propose the intrinsic three-dimensional shape as a new possible diagnostics
to separate classical and disk-like bulges. This is a powerful tool to
unveil the actual nature of galactic bulges and truly address the
demography of classical and disk-like bulges in the nearby universe.

\section{Acknowledgements}
We would like to thank the anonymous referee for the suggestions
that helped us to improve the way we presented our results, 
Dimitri Gadotti for his comments about the photometric and kinematic analysis, and
Alessio Boletti for his remarks about the probabilistic tests.
E.M.C. and L.M. acknowledge financial support
from Padua University through grants DOR1699945, DOR1715817, DOR1885254, and
BIRD164402/16. 
JMA acknowledge support from the Spanish Ministerio de
Economia y Competitividad (MINECO) by the grant AYA2013-43188-P.
L.C. is grateful to the Instituto de Astrof\'isica de Canarias for hospitality while
this paper was in progress.
This research also made use of the NASA/IPAC Extragalactic Database (NED)
which is operated by the Jet Propulsion Laboratory, California
Institute of Technology, under contract with the National Aeronautics
and Space Administration (\url{http://ned.ipac.caltech.edu/}).  We
acknowledge the use of SDSS data (\url{http://www.sdss.org}).


\bsp

\clearpage
\appendix

\section{Additional figures}
\captionsetup{labelfont=bf}
\begin{minipage}{\textwidth}
    \centering
\centering
\includegraphics[width=13.5cm]{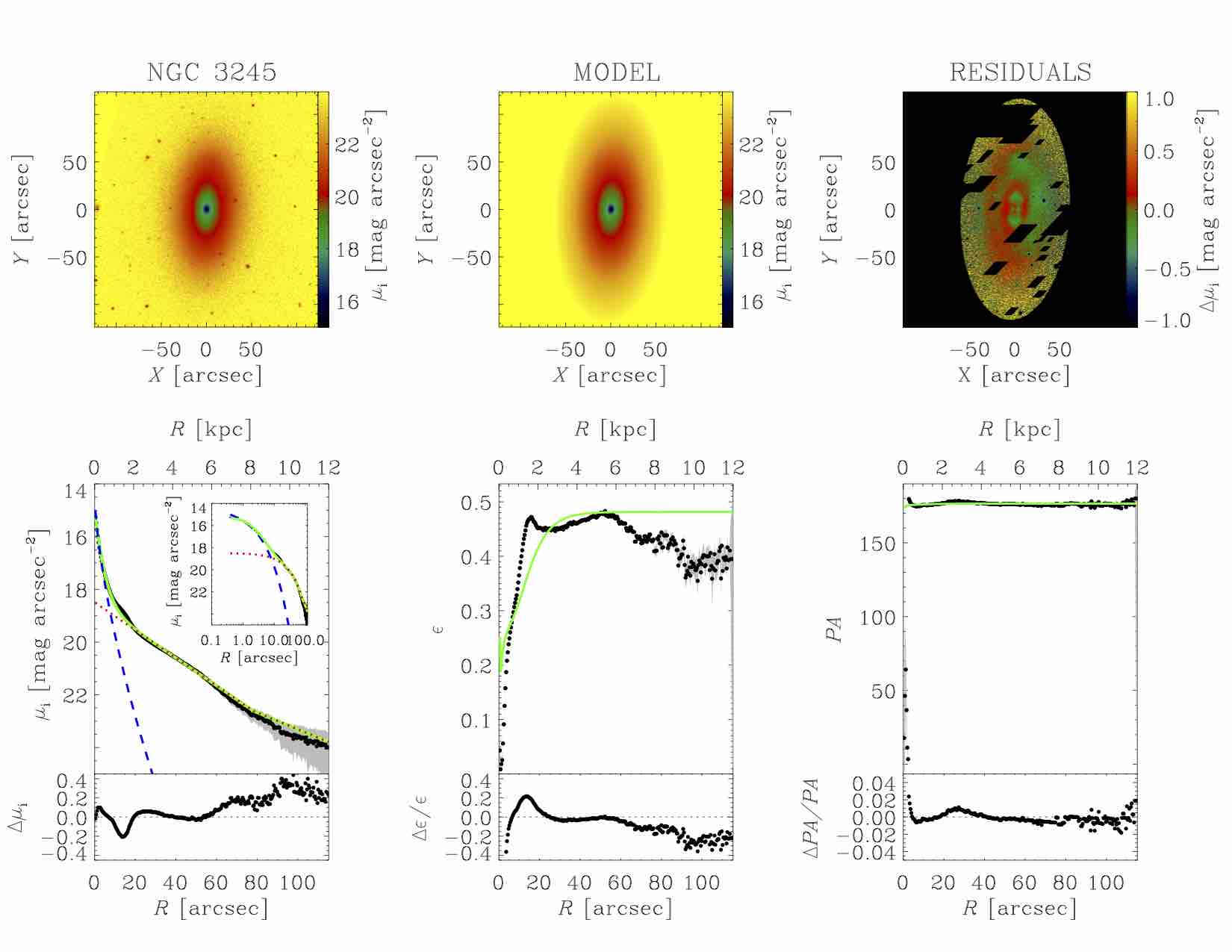}
\includegraphics[width=6.9cm, trim=0.2cm 0.2cm 2.1cm 1cm, clip=true]{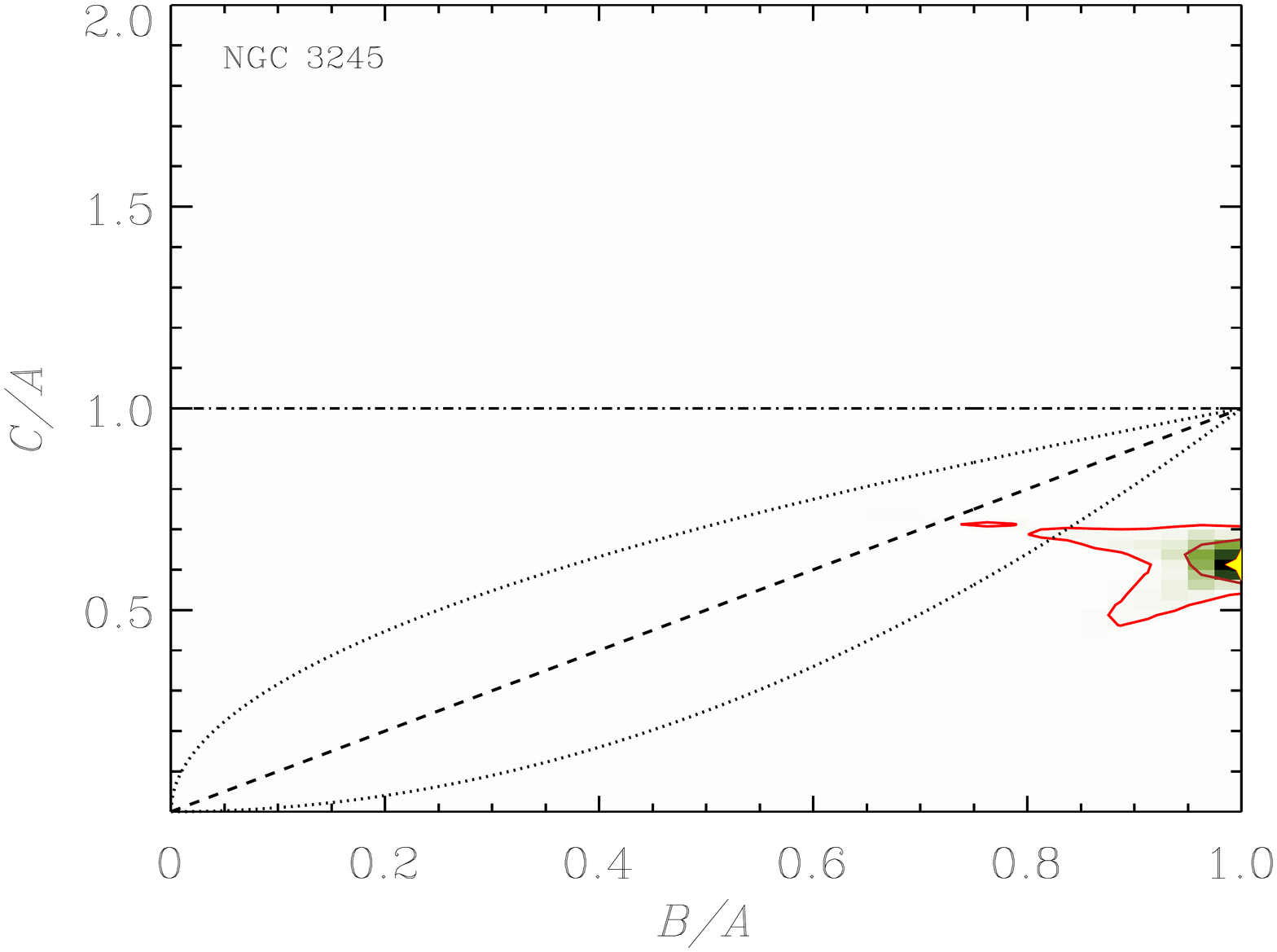}\\
\includegraphics[width=7.7cm, trim=0.5cm 2.4cm 2cm 1.1cm, clip=true]{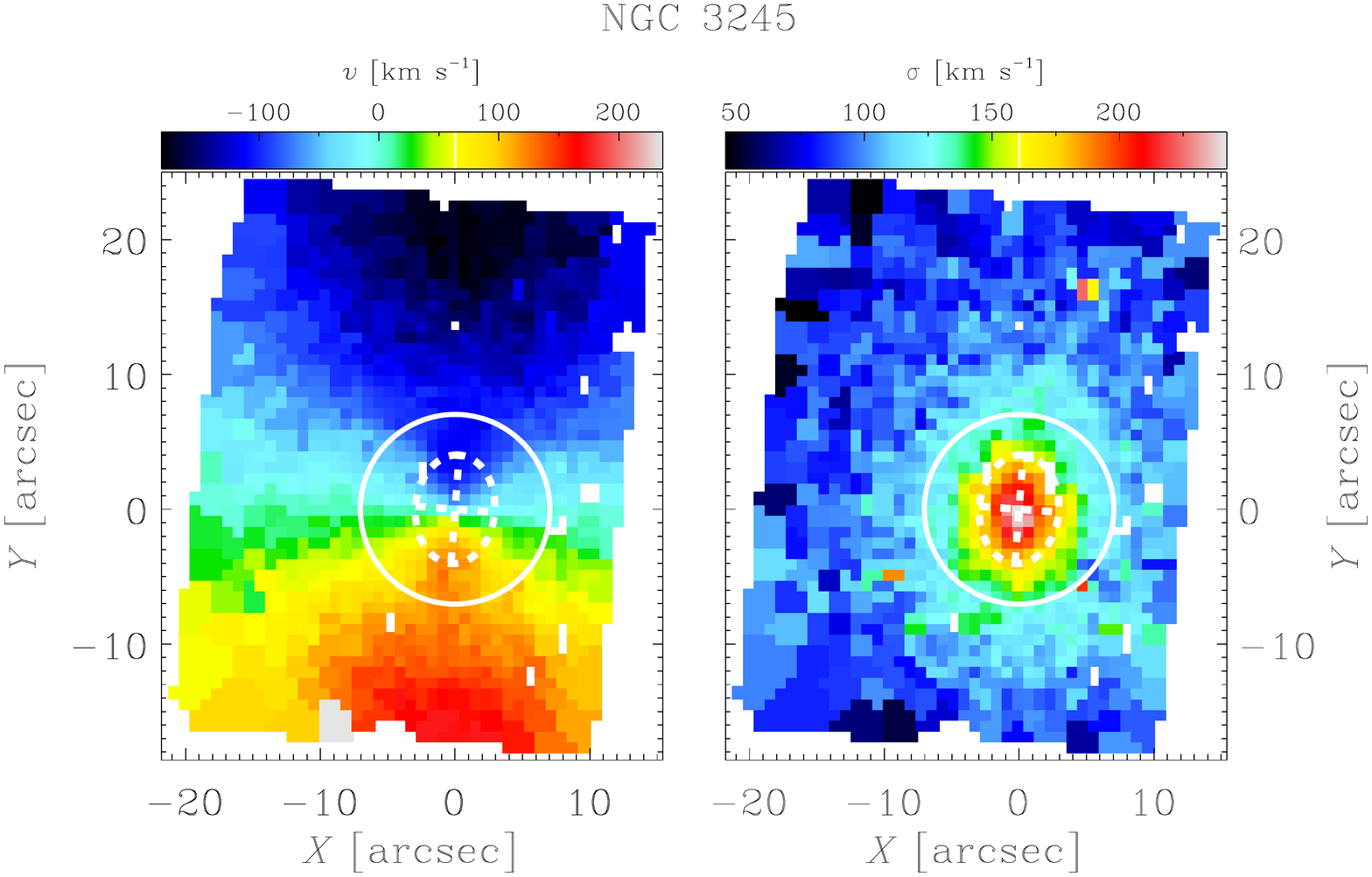}
\includegraphics[width=7.7cm, trim=1.5cm 2.4cm 1cm 0.7cm, clip=true]{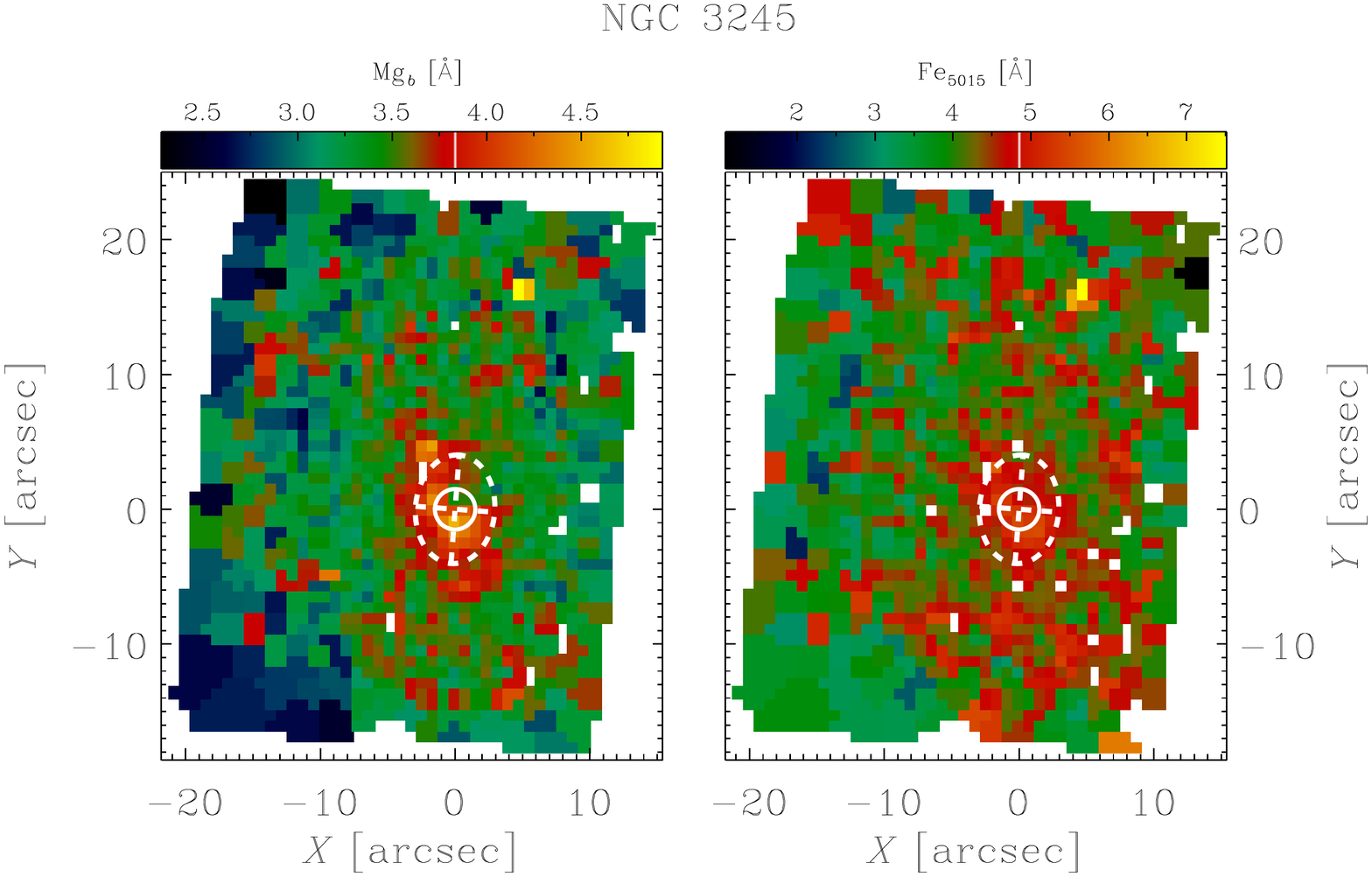}
\captionof{figure}{Two-dimensional photometric decomposition (\emph{top
panels}, as in Fig.~\ref{fig:decomposition_n3156}), distribution
of the intrinsic axial ratios of the bulge (\emph{middle panels}, as
in Fig.~\ref{fig:shape_n3156}), two-dimensional maps of the stellar
kinematics (\emph{bottom left panels}, as in
Fig.~\ref{fig:kinematics_n3156}) and line-strength indices
(\emph{bottom right panels}, as in Fig.~\ref{fig:indices_n3156}) of
the sample galaxies, except for NGC~3156. The galaxy name is given
in each plot.}
\label{fig:appendix}
\end{minipage}

\begin{figure*}
\centering
\addtocounter{figure}{-1}
\includegraphics[width=13.5cm]{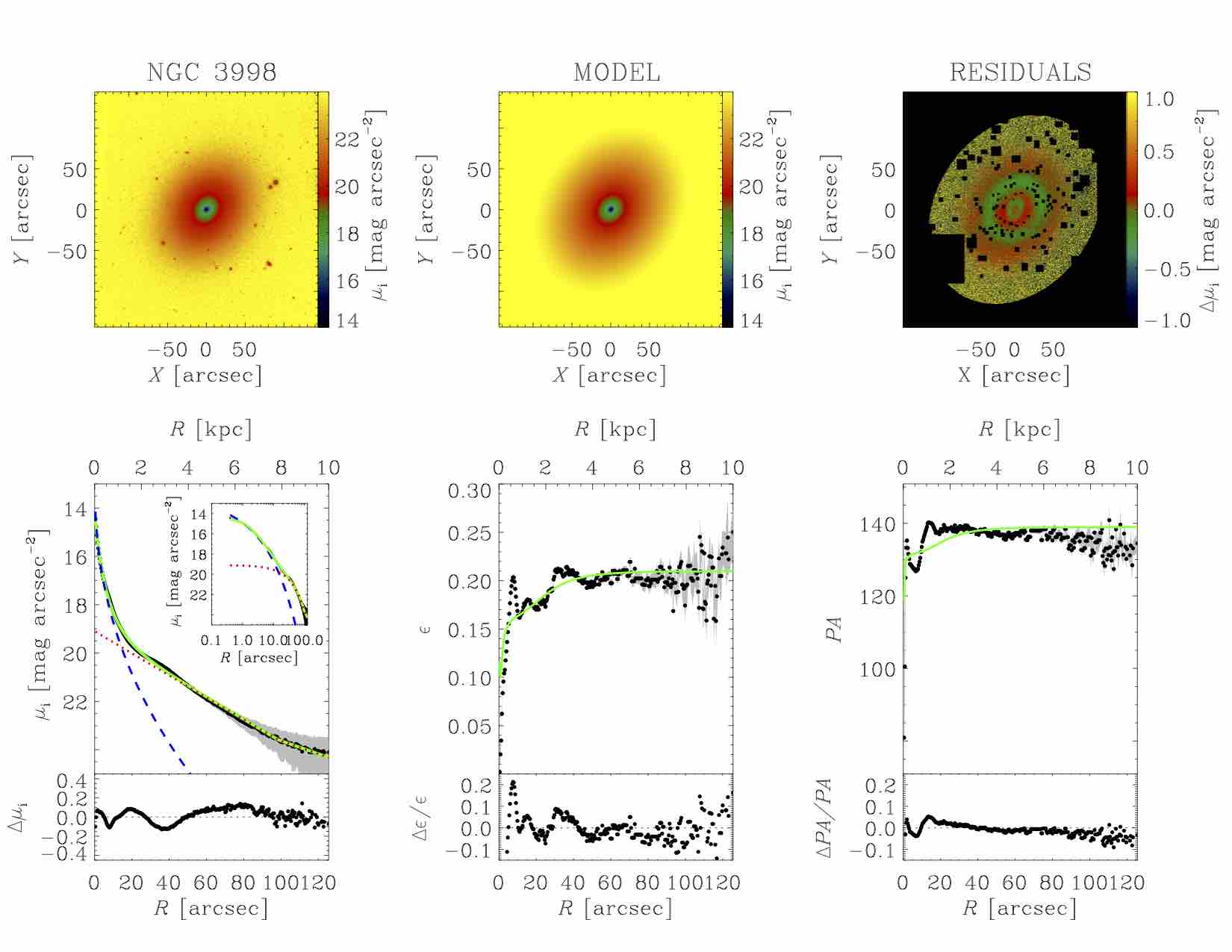}
\includegraphics[width=6.9cm, trim=0.2cm 0.2cm 2.1cm 1cm, clip=true]{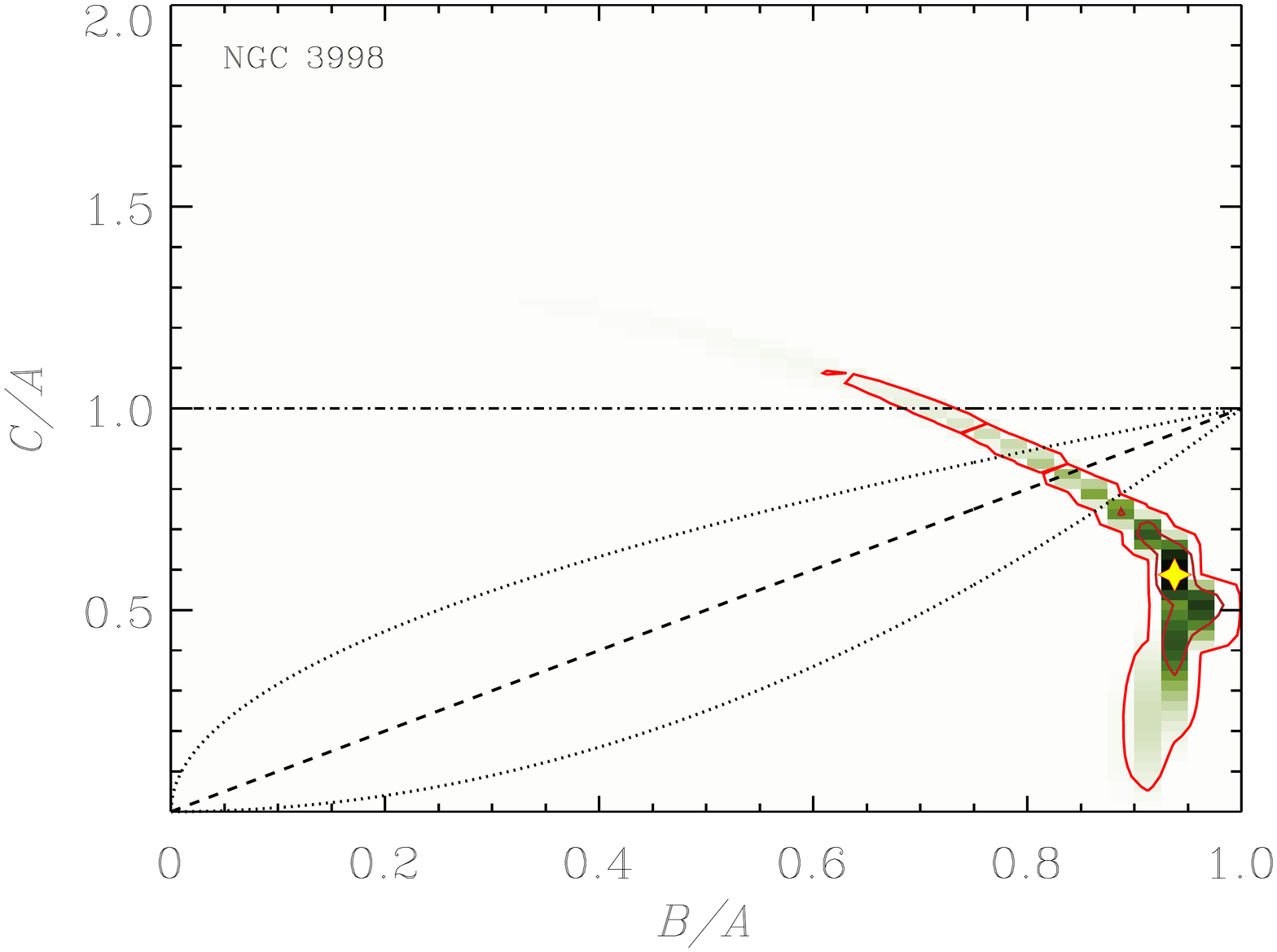}\\
\includegraphics[width=7.7cm, trim=0.5cm 2.4cm 2cm 1.1cm, clip=true]{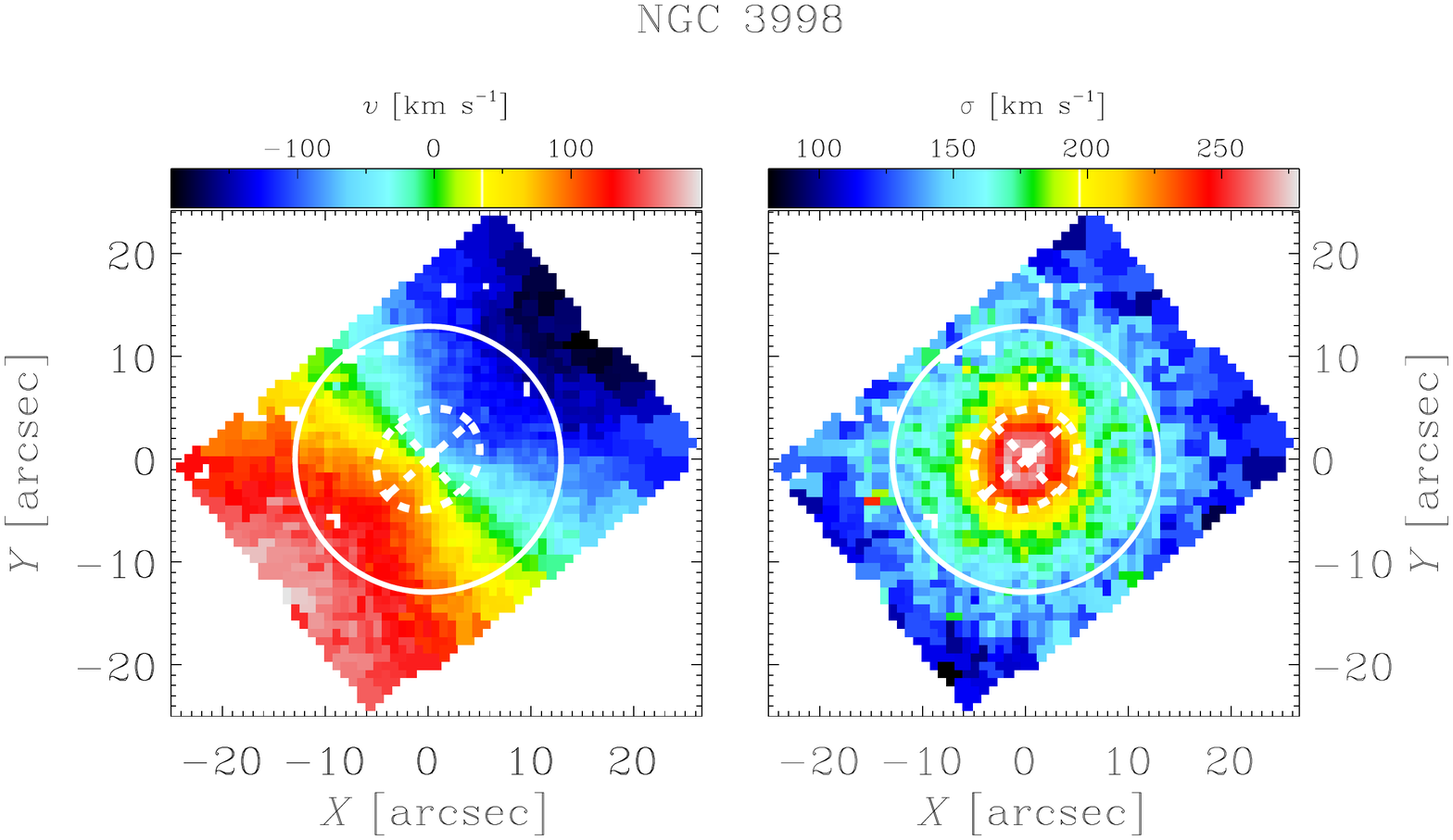}
\includegraphics[width=7.7cm, trim=1.5cm 2.4cm 1cm 0.7cm, clip=true]{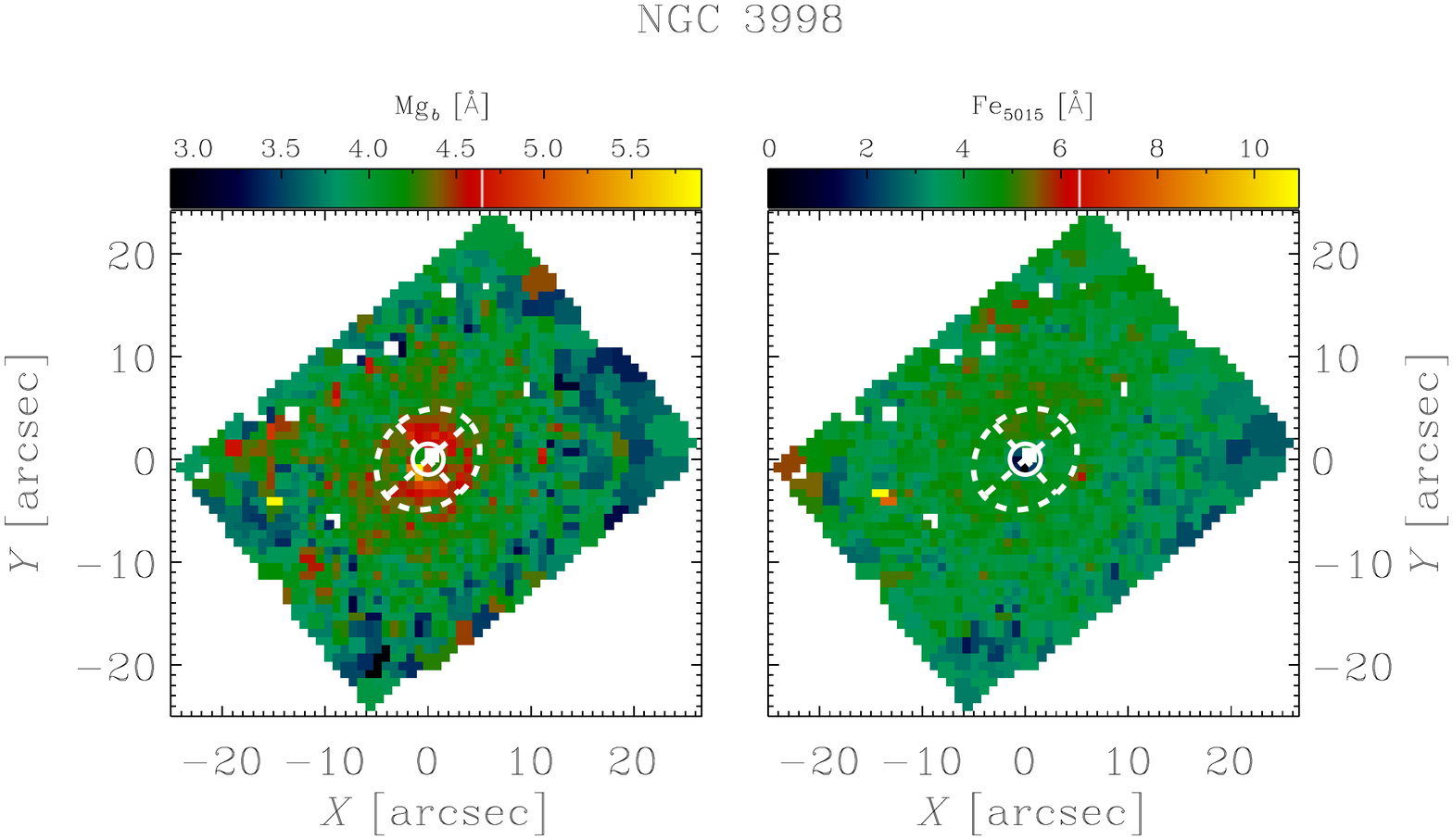}
\caption{continued}
\end{figure*}

\begin{figure*}
\centering
\addtocounter{figure}{-1}
\includegraphics[width=13.5cm]{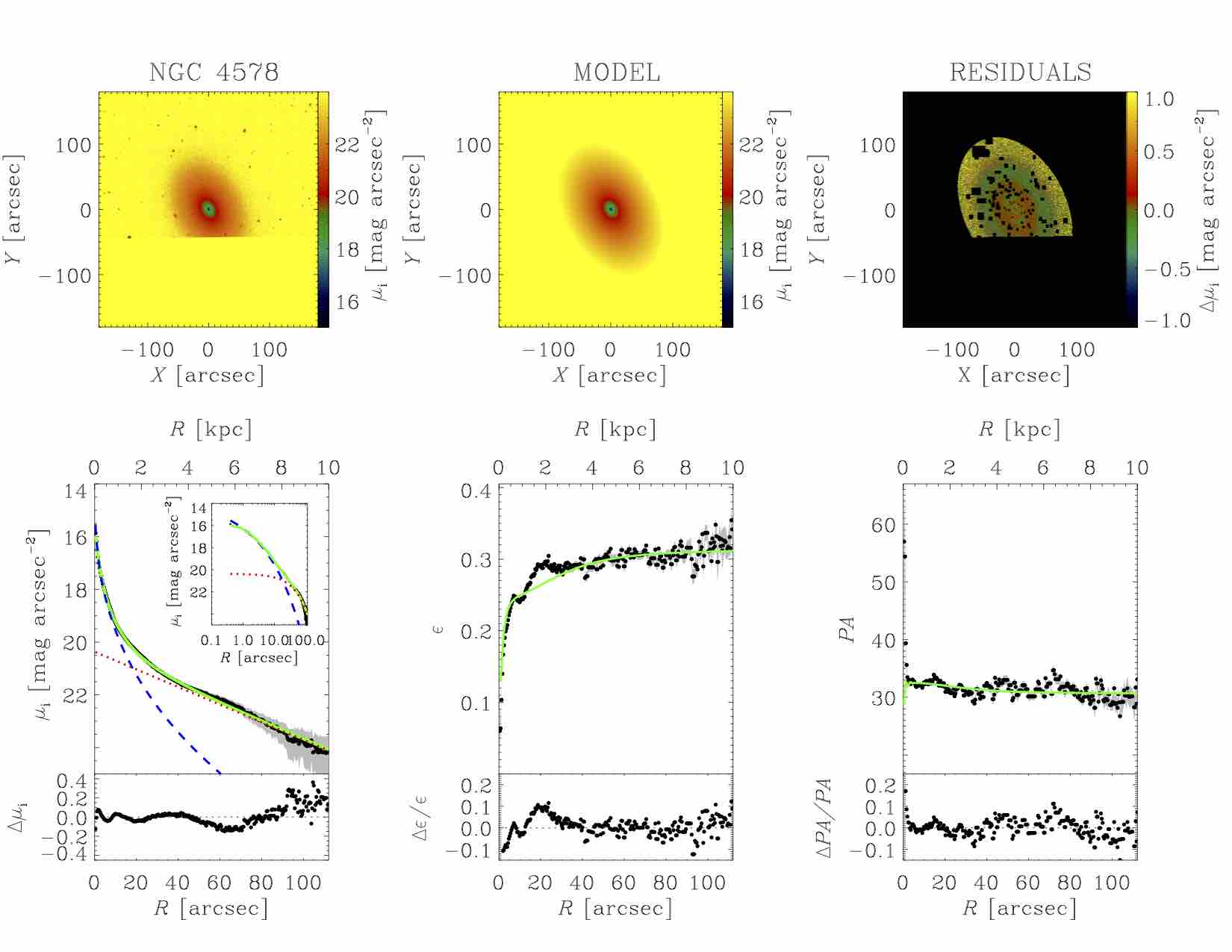}
\includegraphics[width=6.9cm, trim=0.2cm 0.2cm 2.1cm 1cm, clip=true]{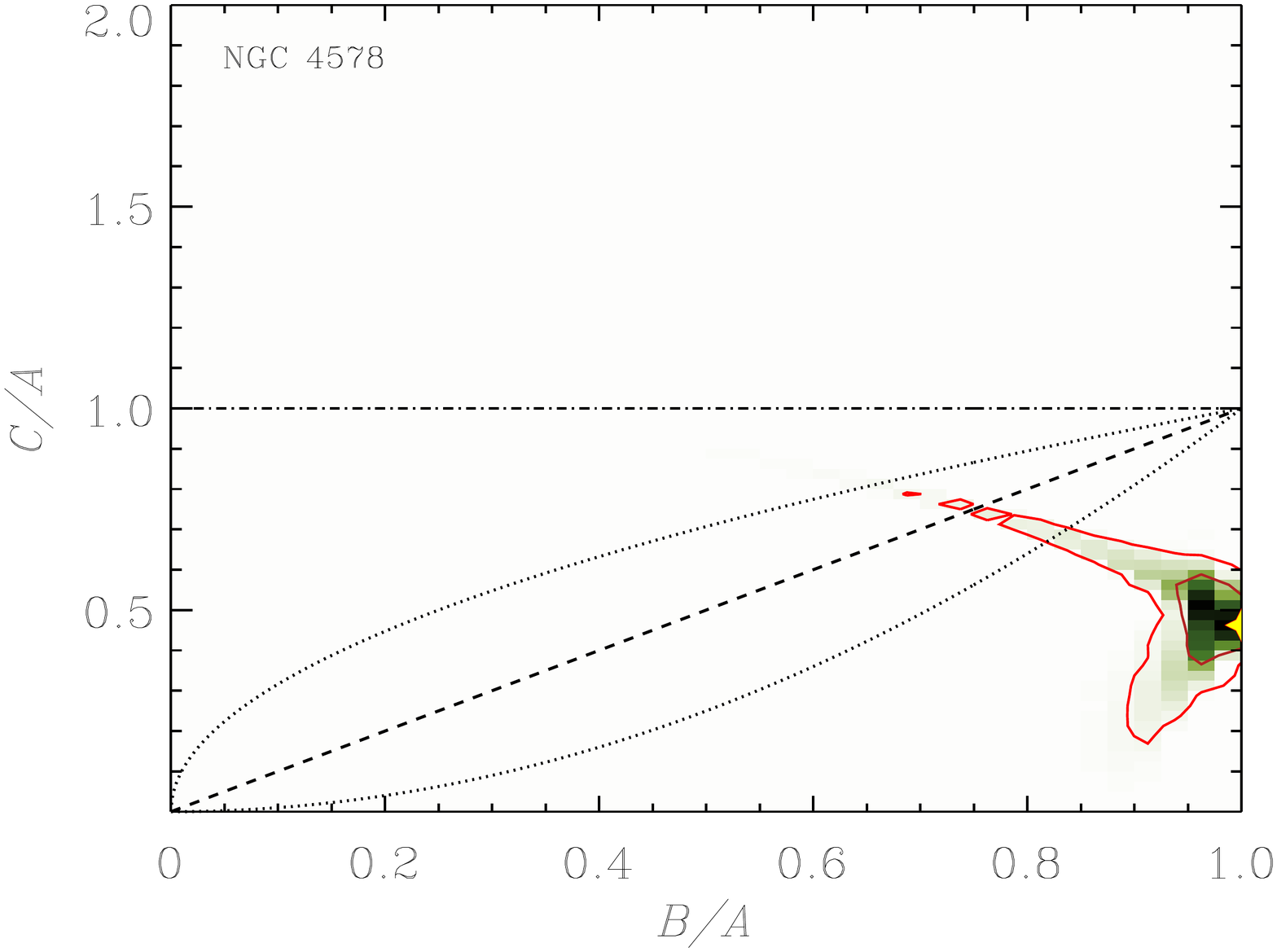}\\
\includegraphics[width=7.7cm, trim=0.5cm 2.4cm 2cm 1.1cm, clip=true]{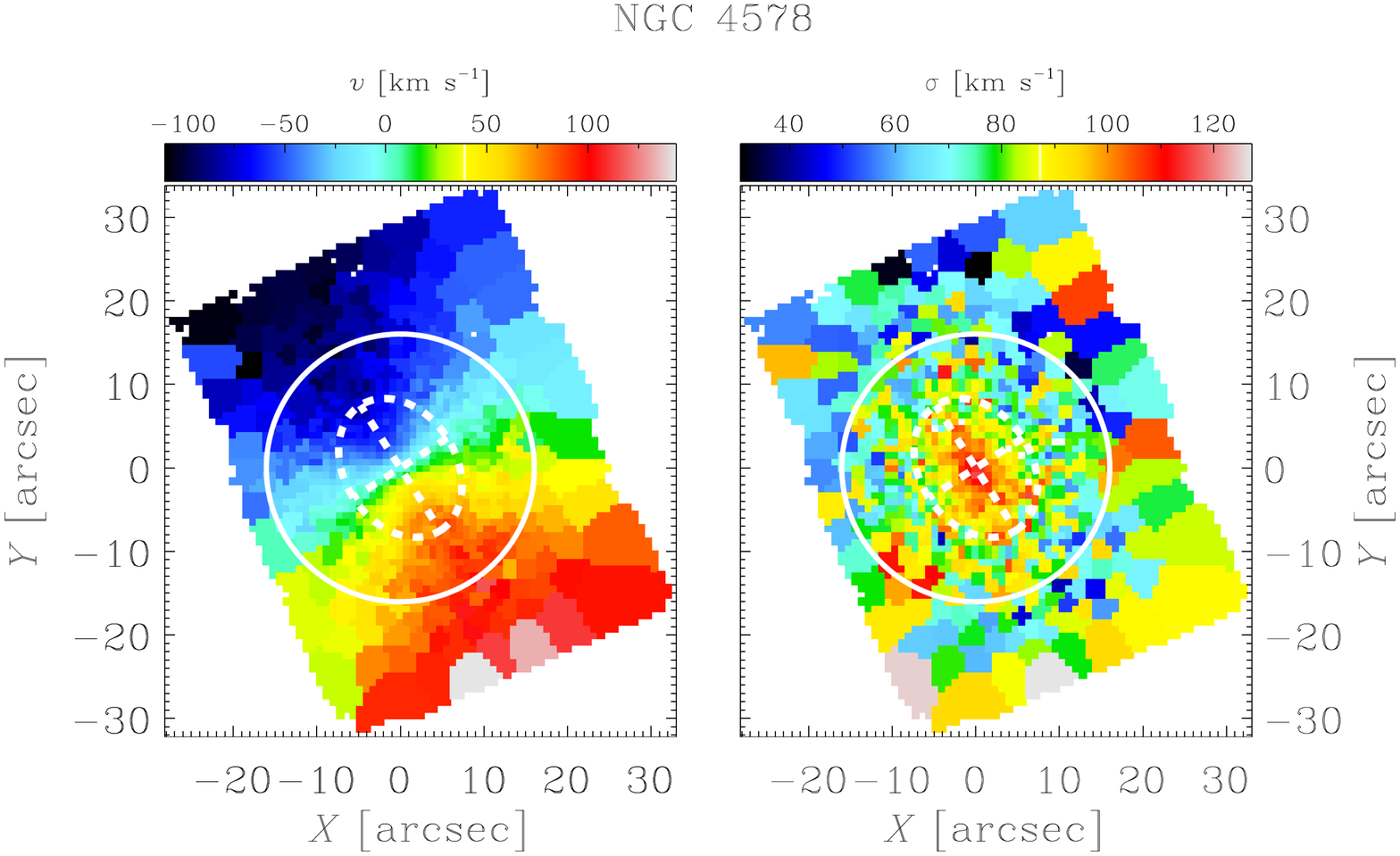}
\includegraphics[width=7.7cm, trim=1.5cm 2.4cm 1cm 0.7cm, clip=true]{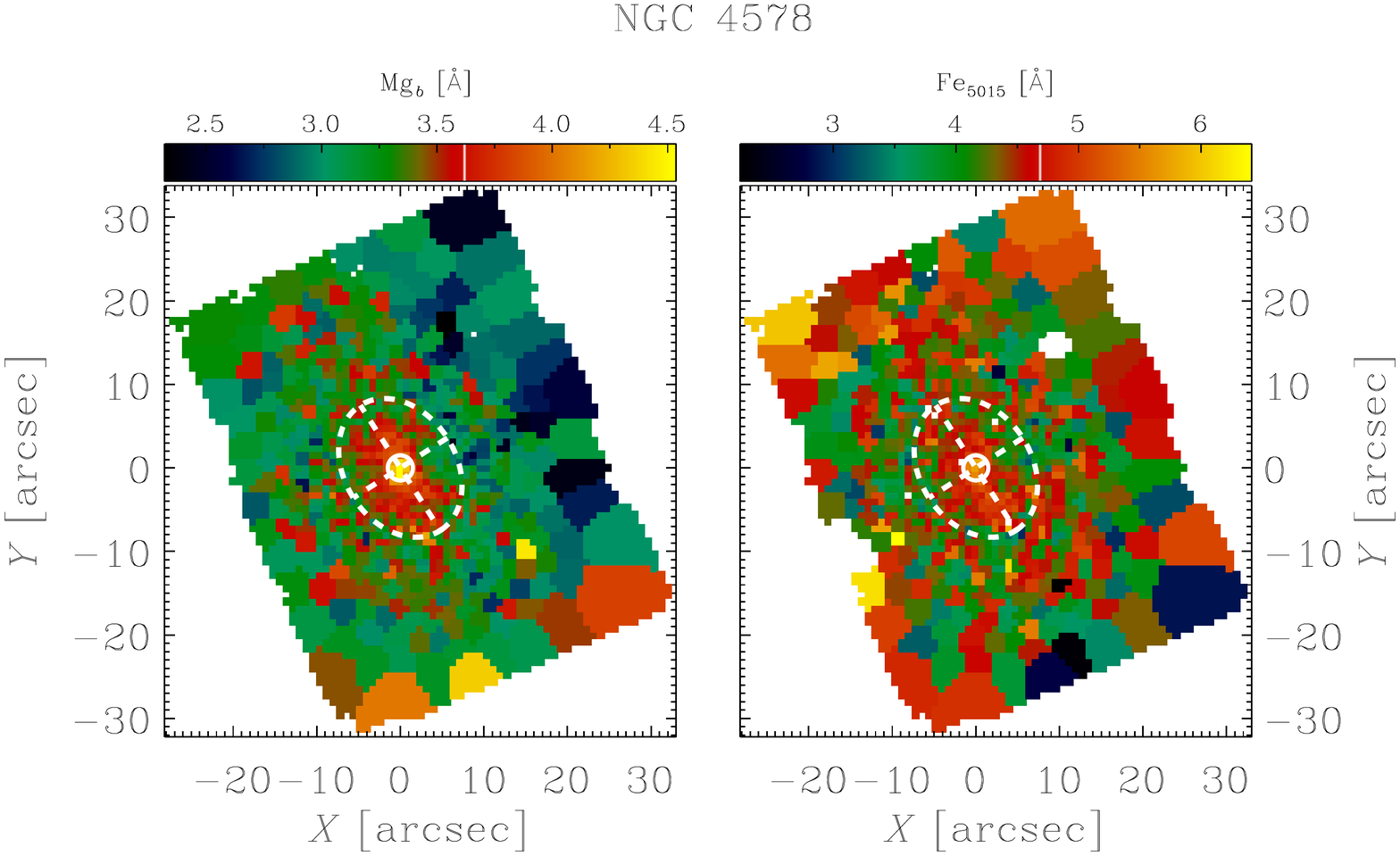}
\caption{continued.}
\end{figure*}

\begin{figure*}
\centering
\addtocounter{figure}{-1}
\includegraphics[width=13.5cm]{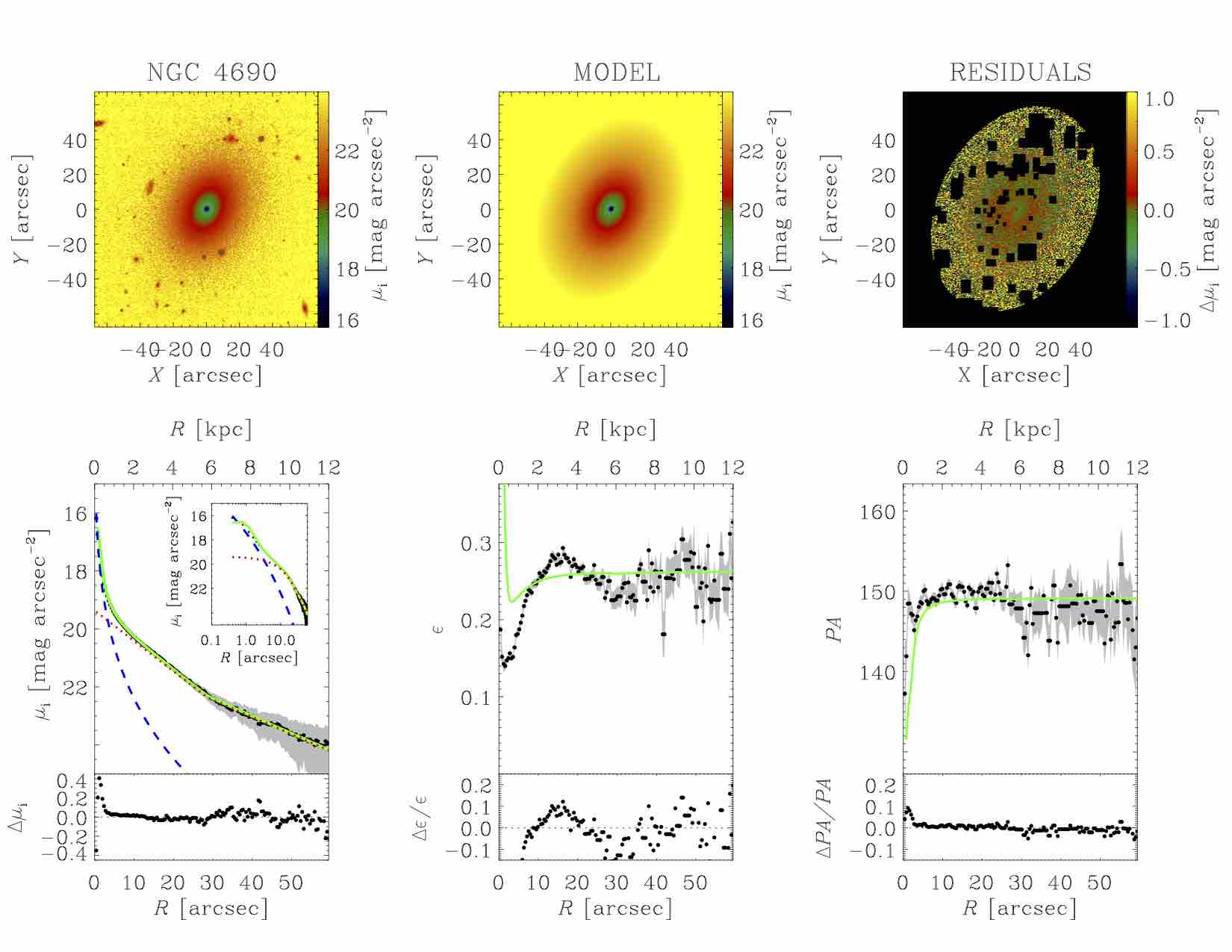}
\includegraphics[width=6.9cm, trim=0.2cm 0.2cm 2.1cm 1cm, clip=true]{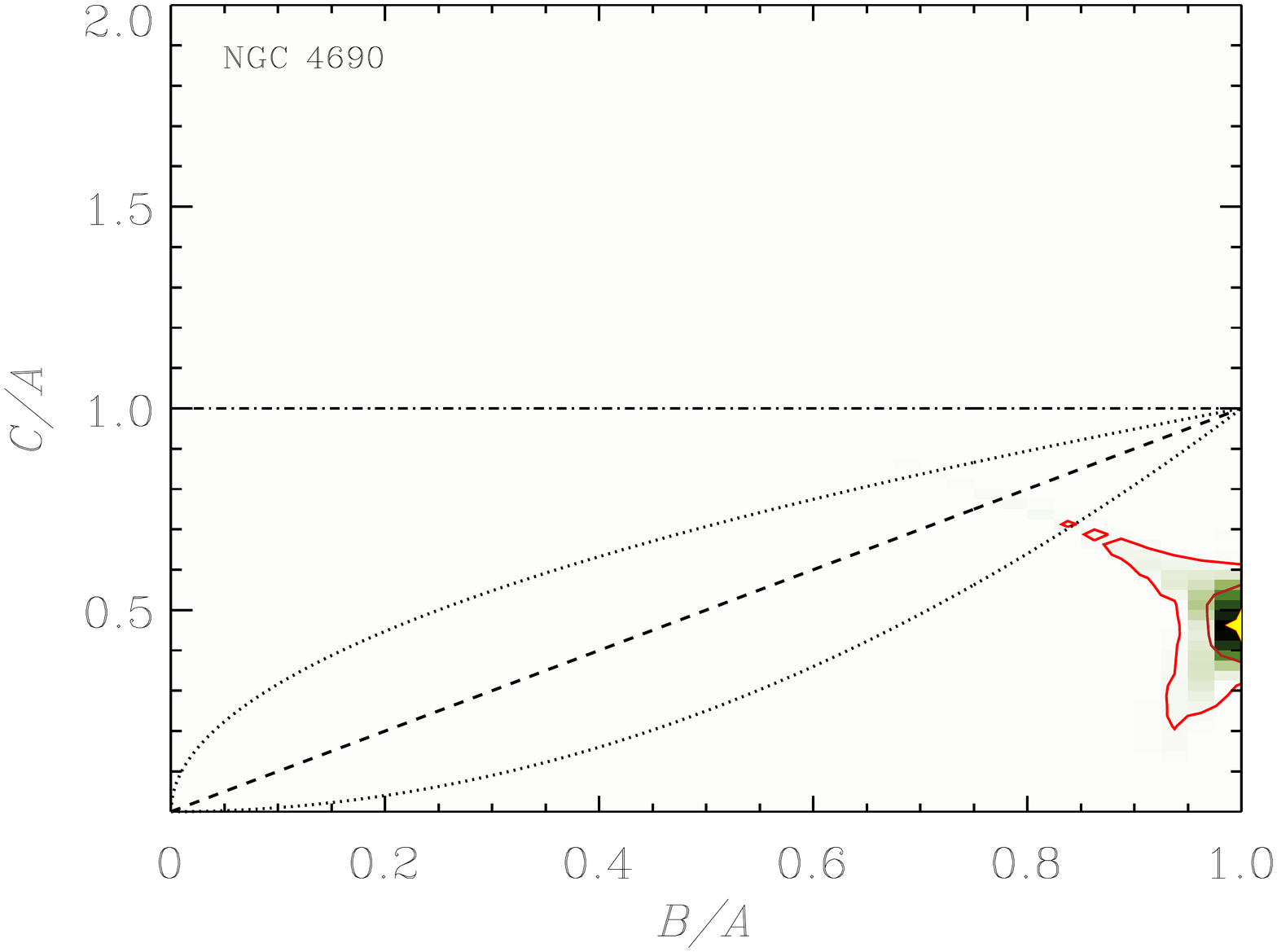}\\
\includegraphics[width=7.7cm, trim=0.5cm 2.4cm 2cm 1.1cm, clip=true]{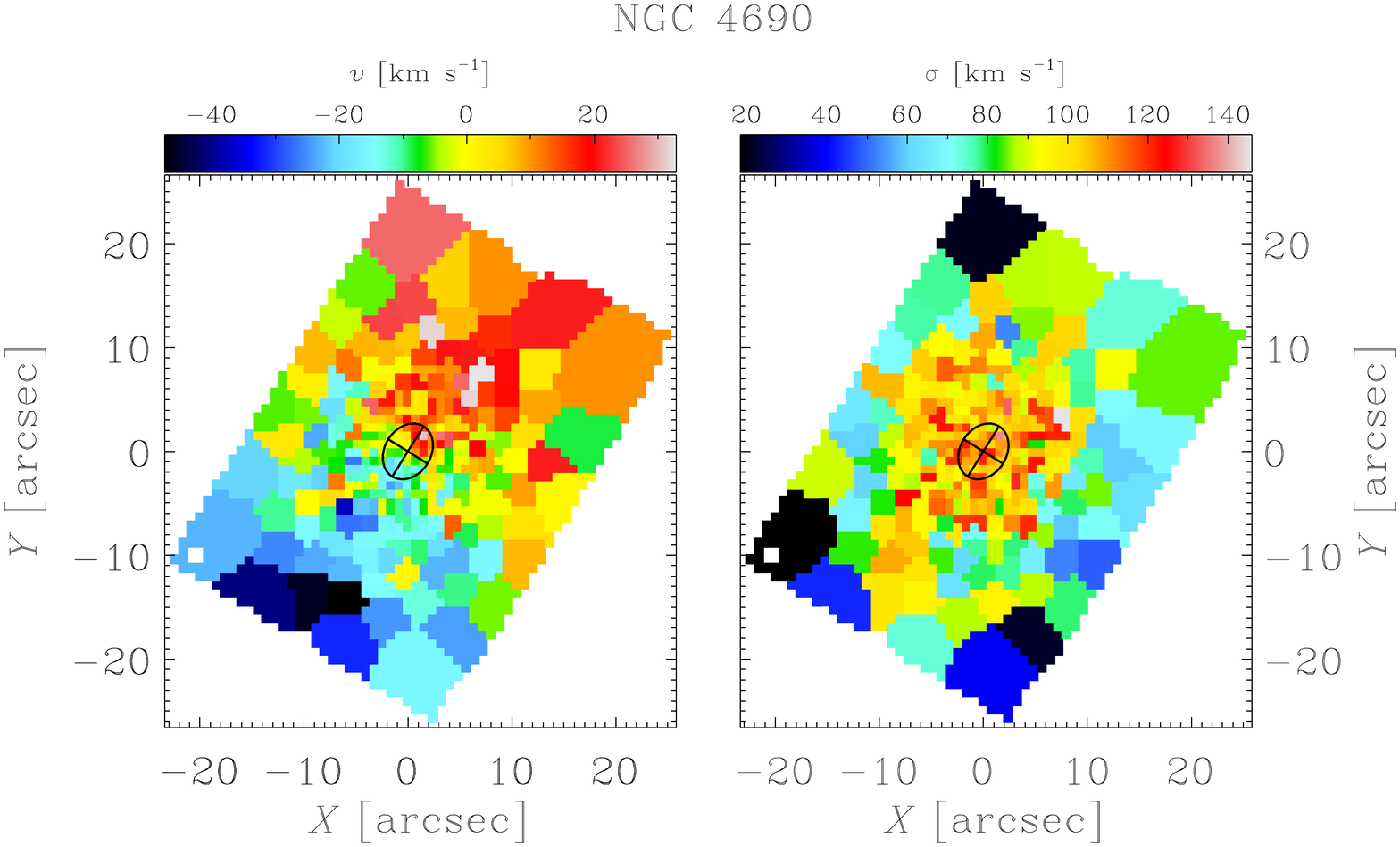}
\includegraphics[width=7.7cm, trim=1.5cm 2.4cm 1cm 0.7cm, clip=true]{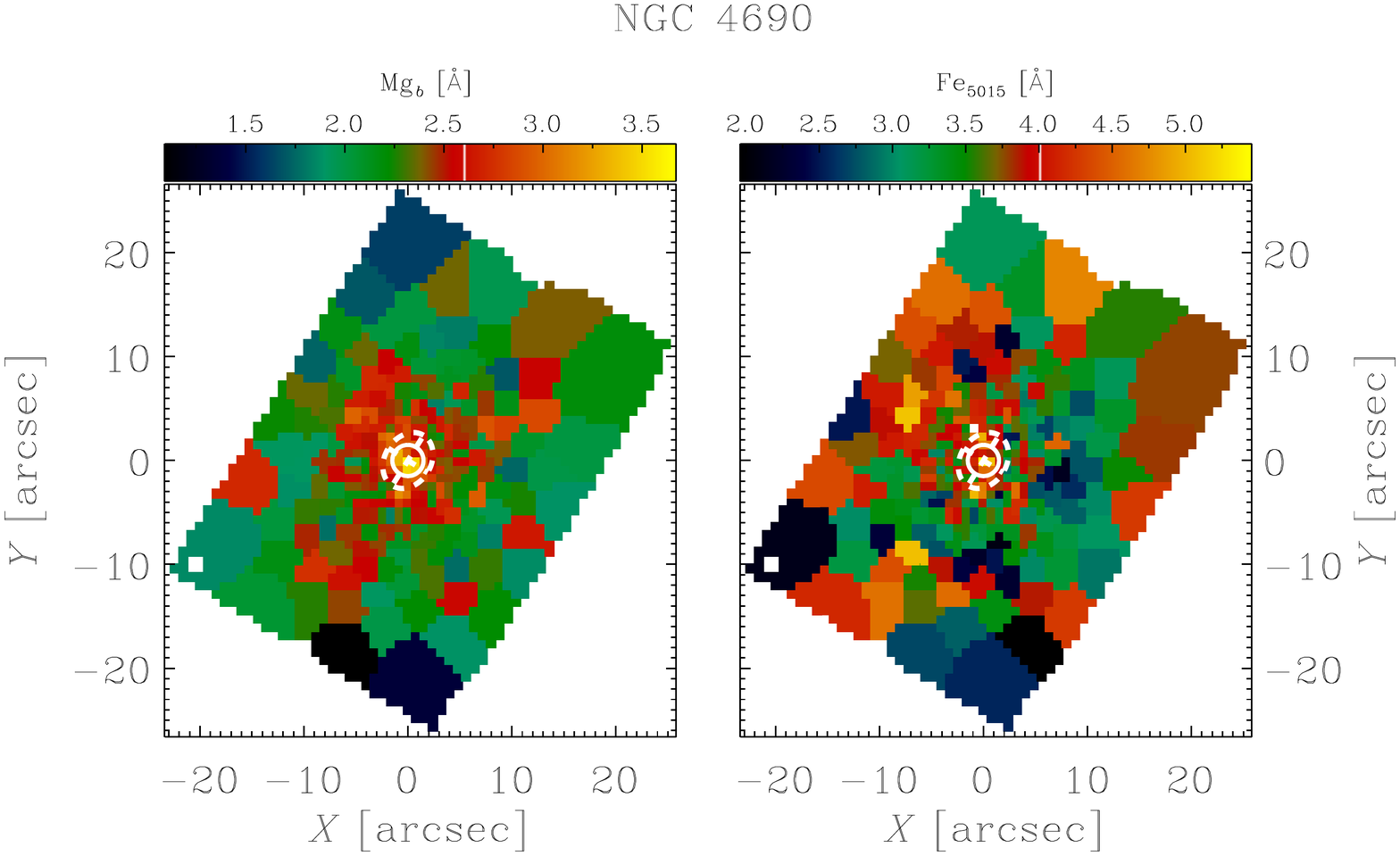}
\caption{continued.}
\end{figure*}

\begin{figure*}
\centering
\addtocounter{figure}{-1}
\includegraphics[width=13.5cm]{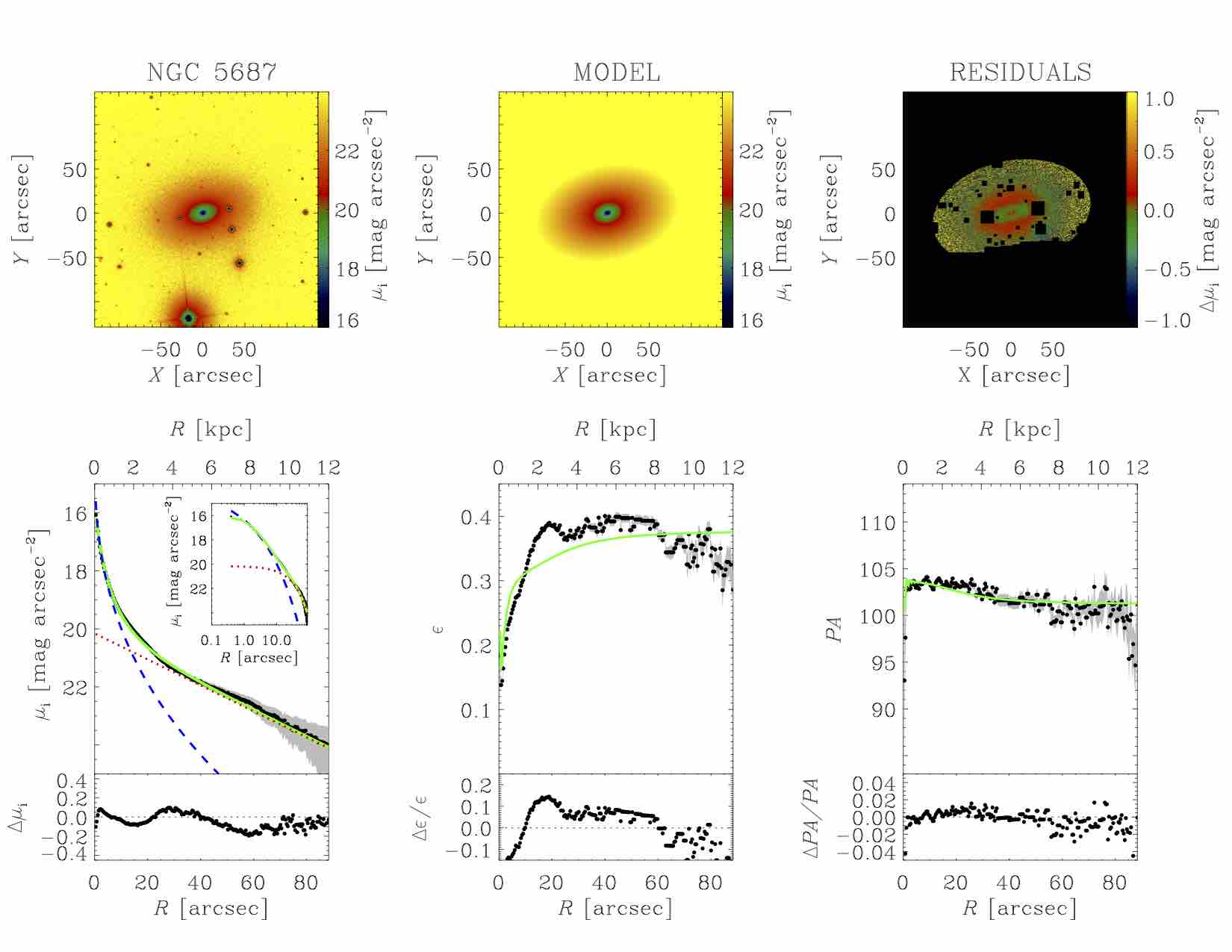}
\includegraphics[width=6.9cm, trim=0.2cm 0.2cm 2.1cm 1cm, clip=true]{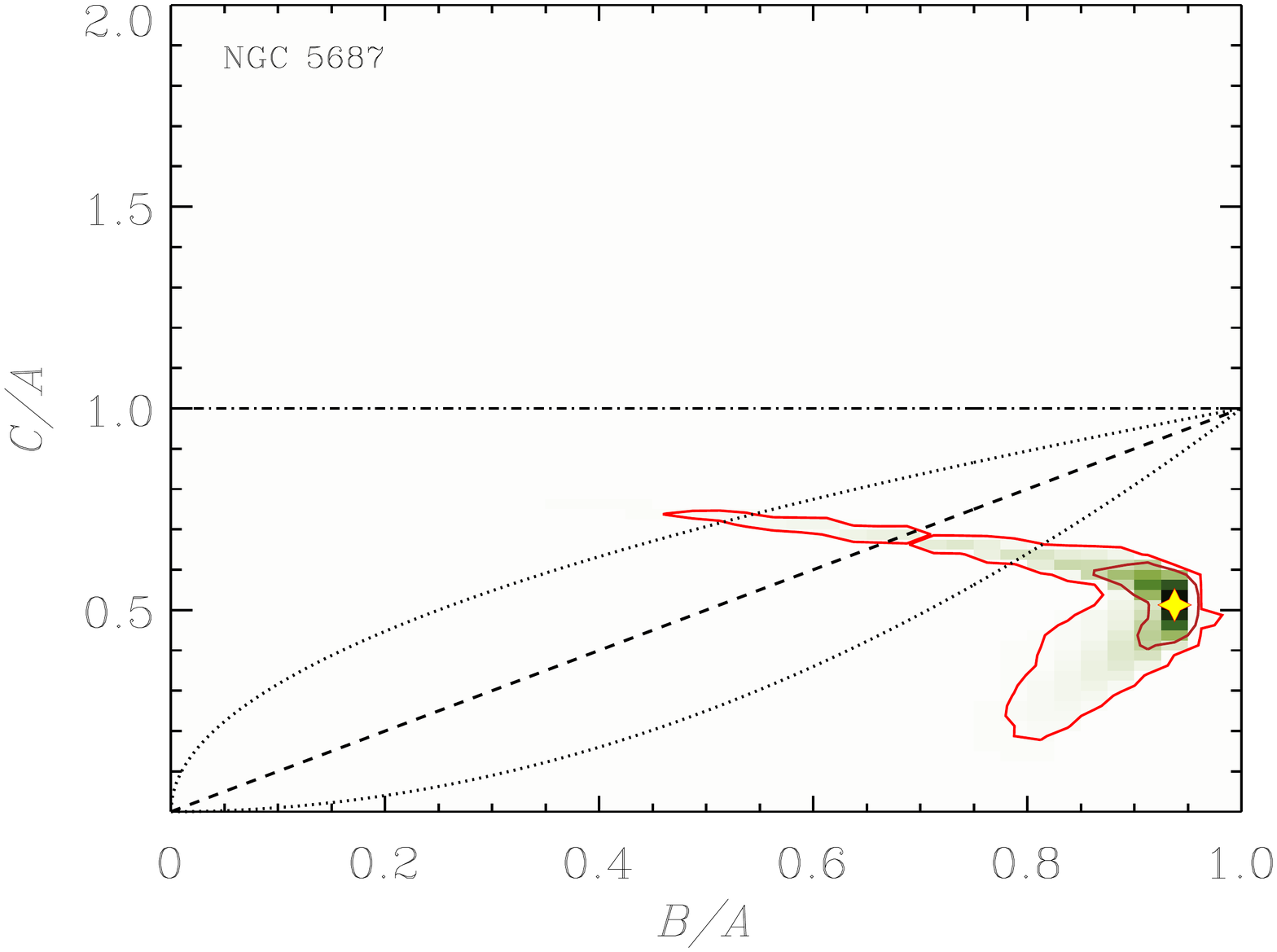}\\
\includegraphics[width=7.7cm, trim=0.5cm 2.4cm 2cm 1.1cm, clip=true]{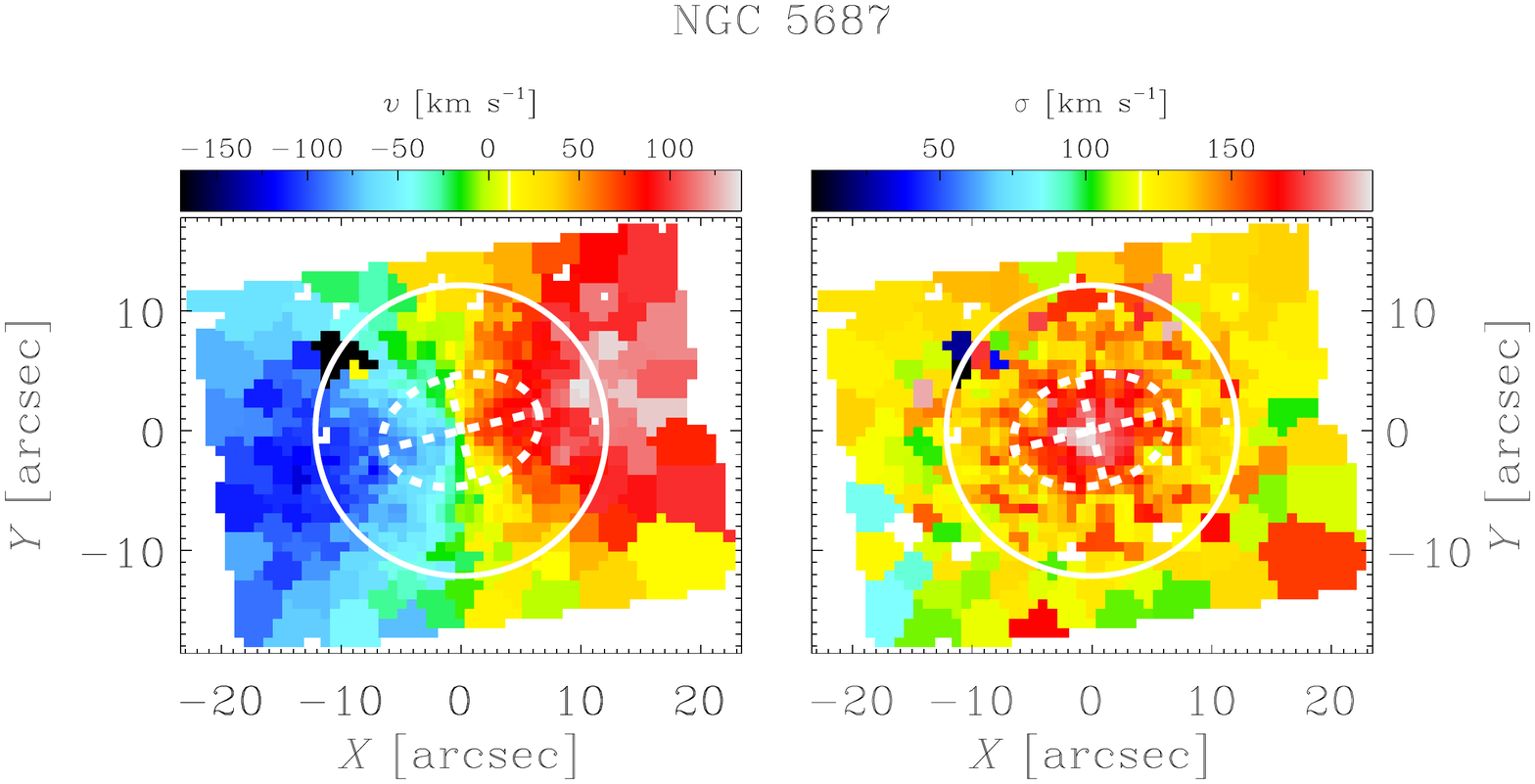}
\includegraphics[width=7.7cm, trim=1.5cm 2.4cm 1cm 0.7cm, clip=true]{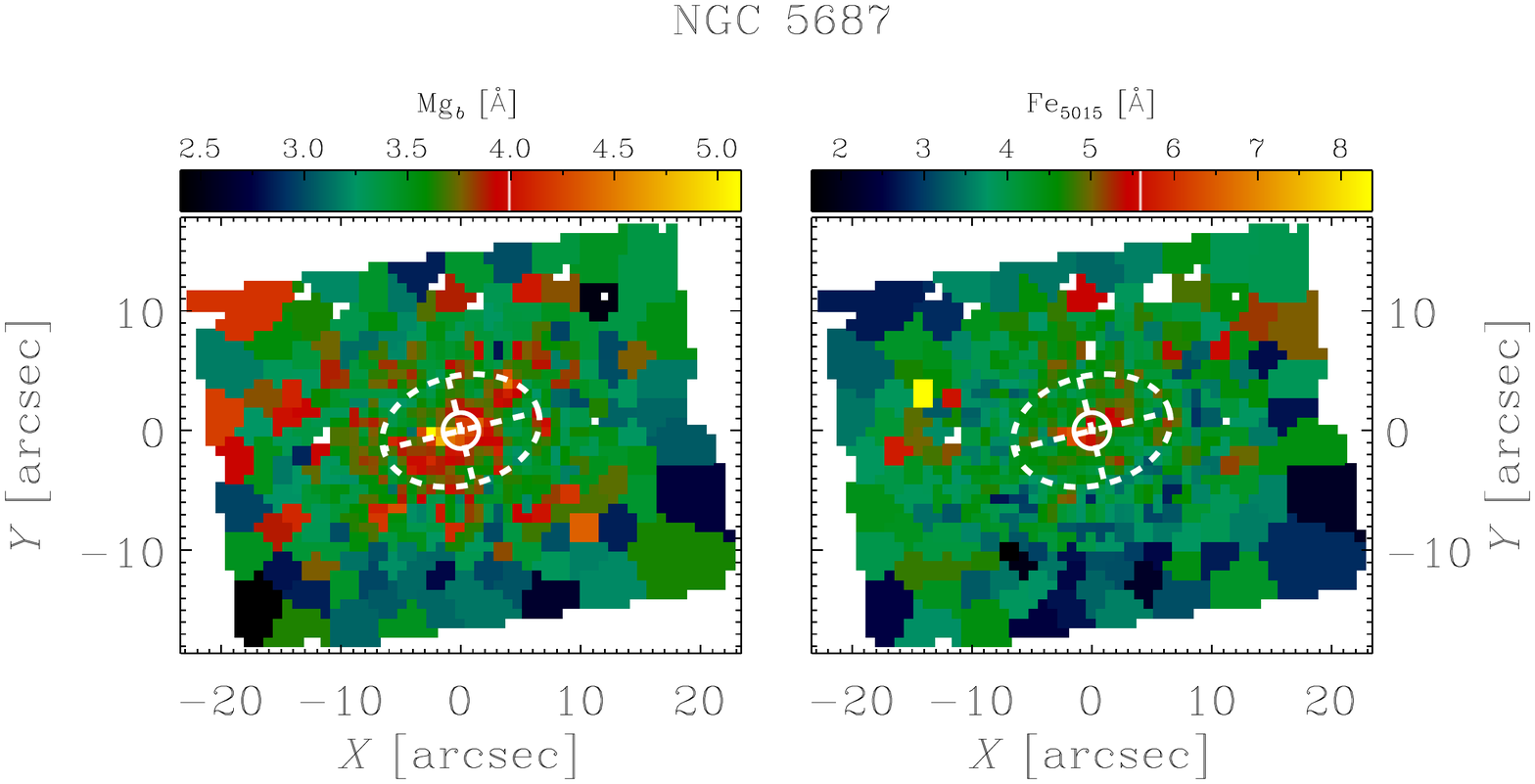}
\caption{continued.}
\end{figure*}

\begin{figure*}
\centering
\addtocounter{figure}{-1}
\includegraphics[width=13.5cm]{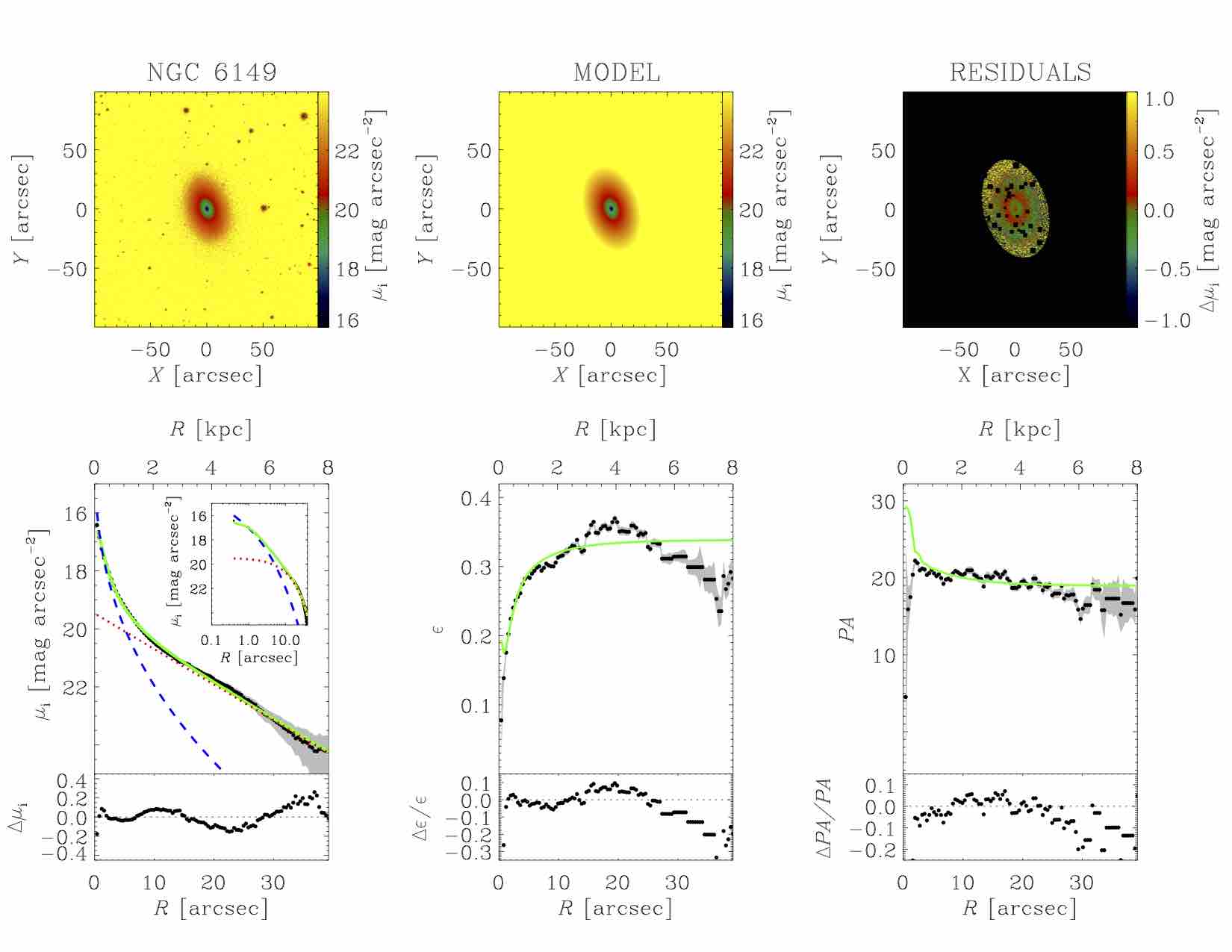}
\includegraphics[width=6.9cm, trim=0.2cm 0.2cm 2.1cm 1cm, clip=true]{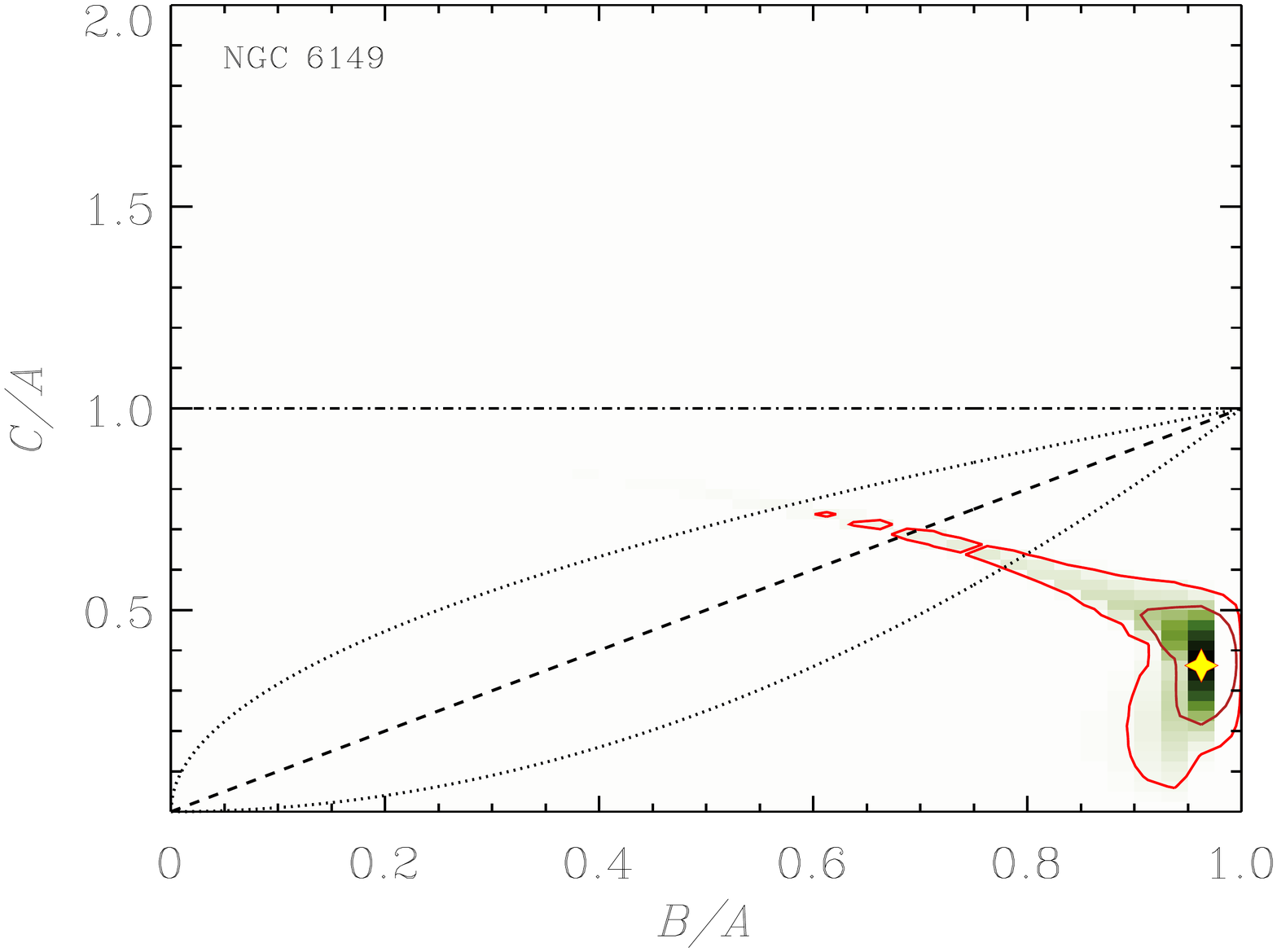}\\
\includegraphics[width=7.7cm, trim=0.5cm 2.4cm 2cm 1.1cm, clip=true]{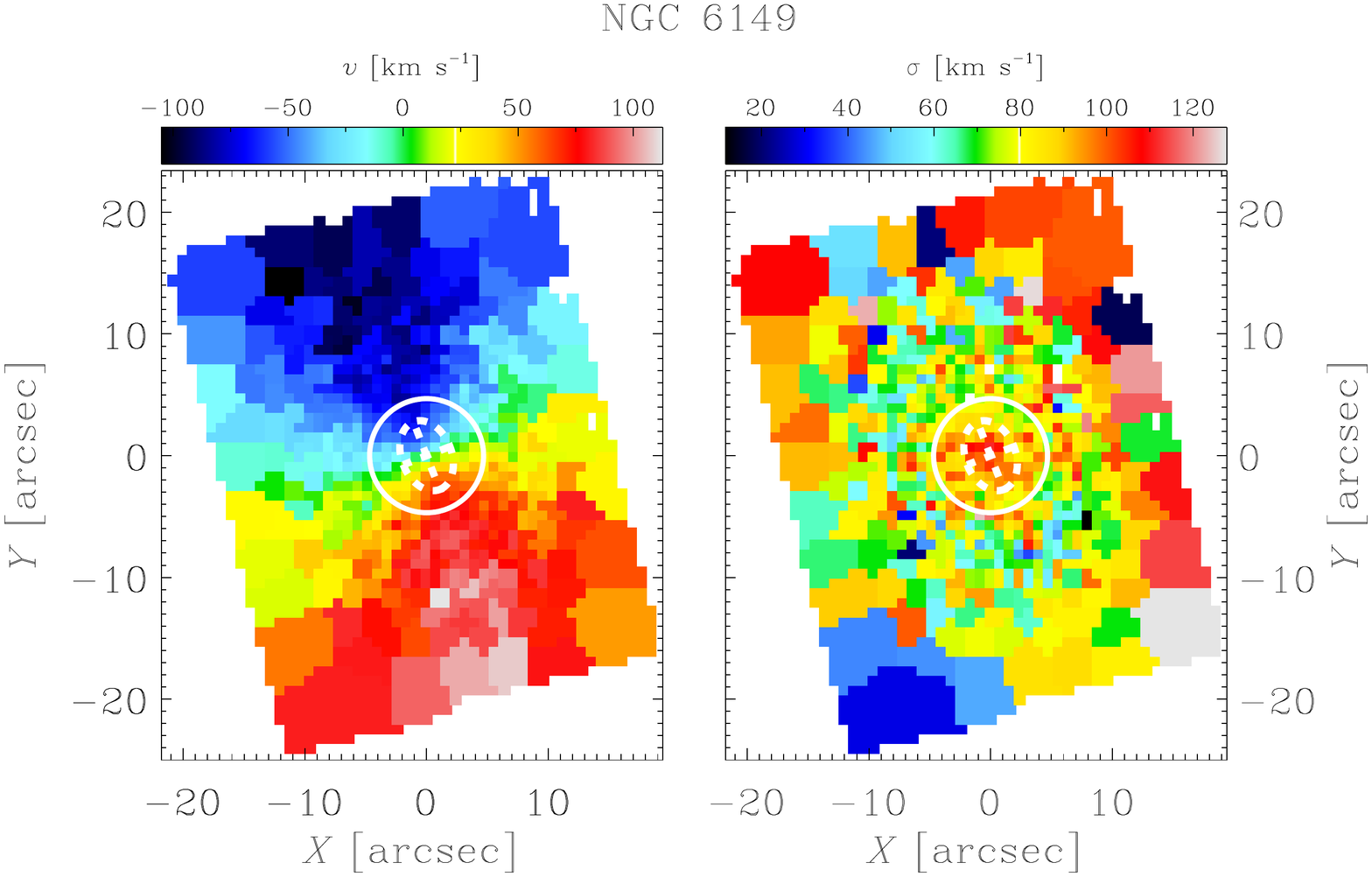}
\includegraphics[width=7.7cm, trim=1.5cm 2.4cm 1cm 0.7cm, clip=true]{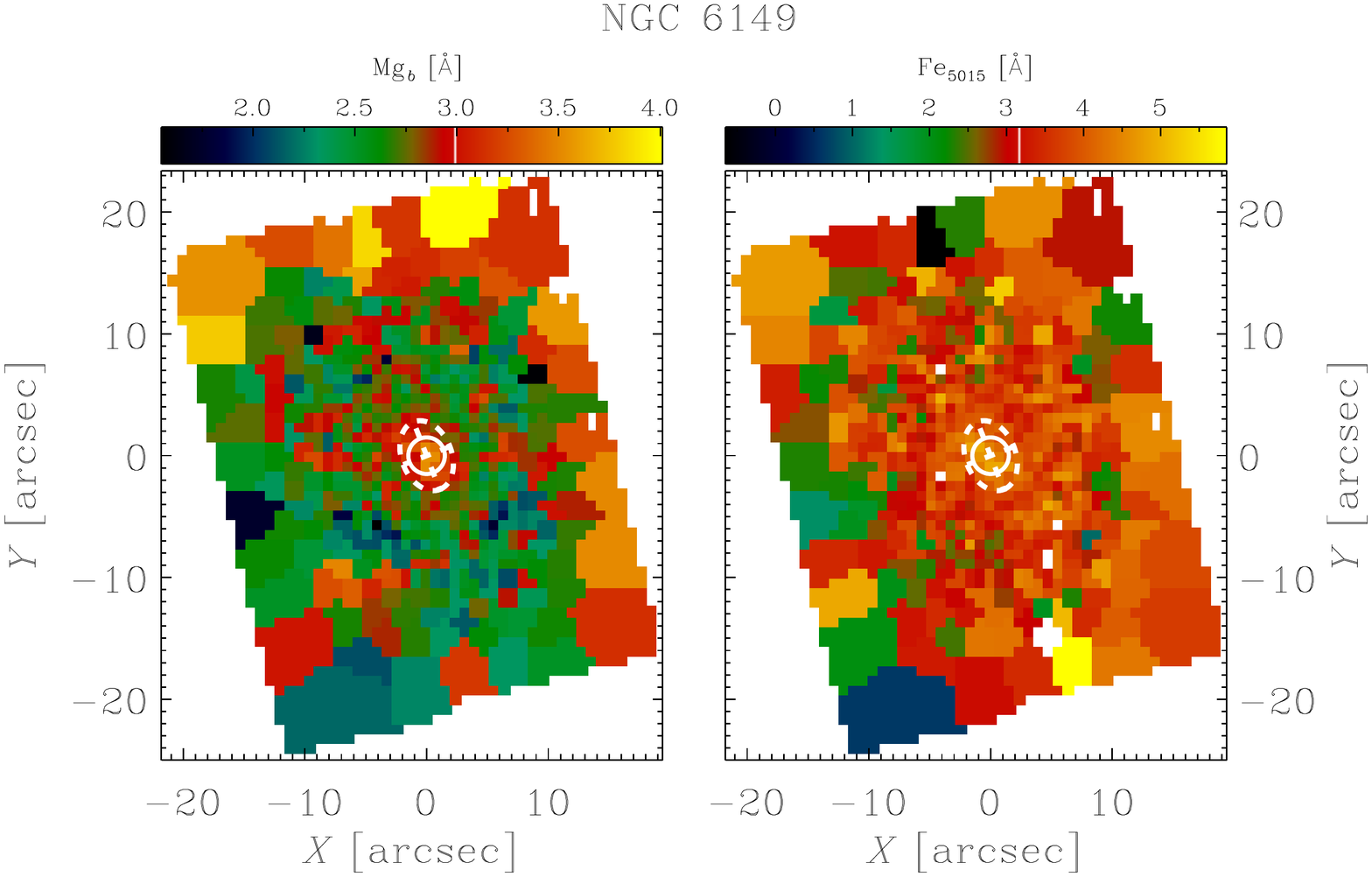}
\caption{continued.}
\end{figure*}

\begin{figure*}
\centering
\addtocounter{figure}{-1}
\includegraphics[width=13.5cm]{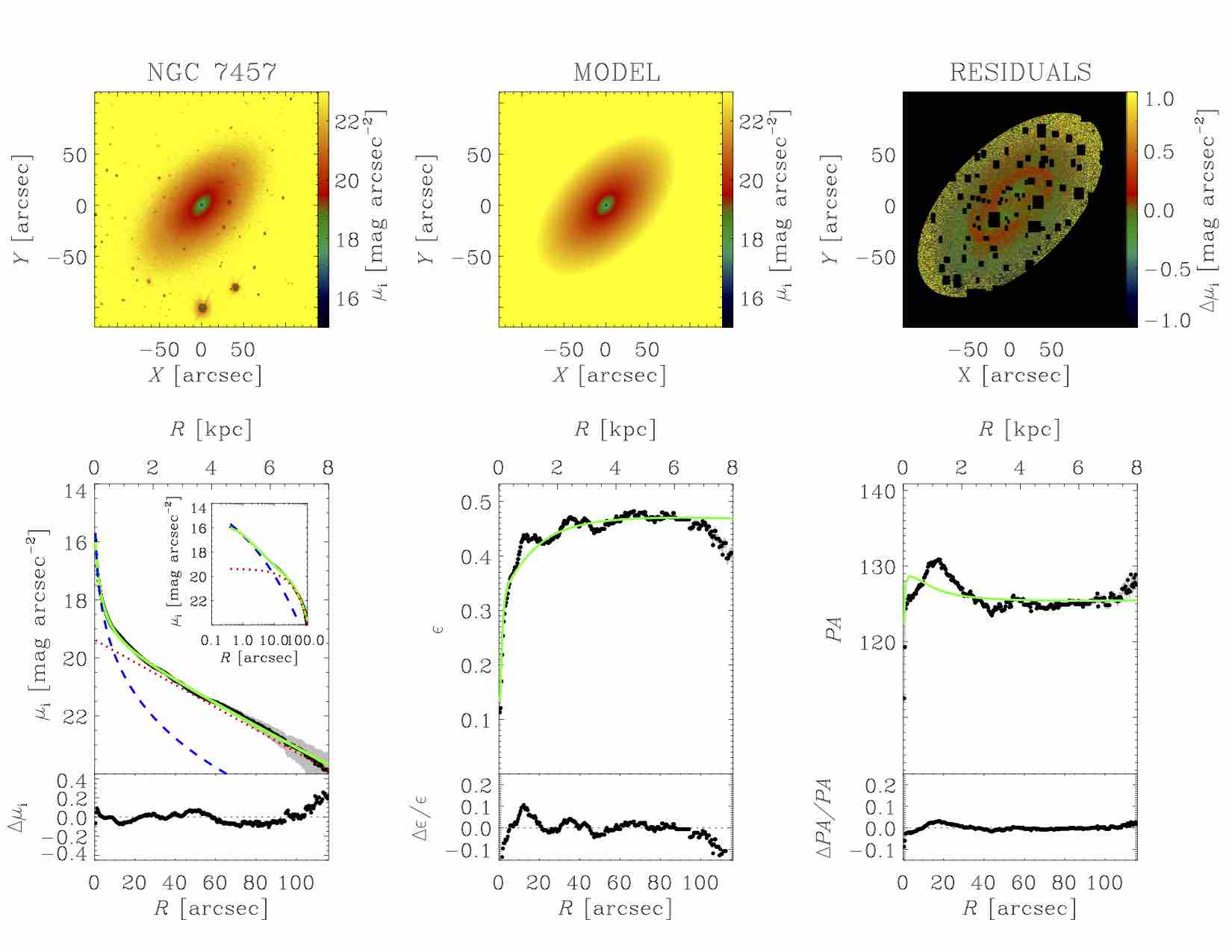}
\includegraphics[width=6.9cm, trim=0.2cm 0.2cm 2.1cm 1cm, clip=true]{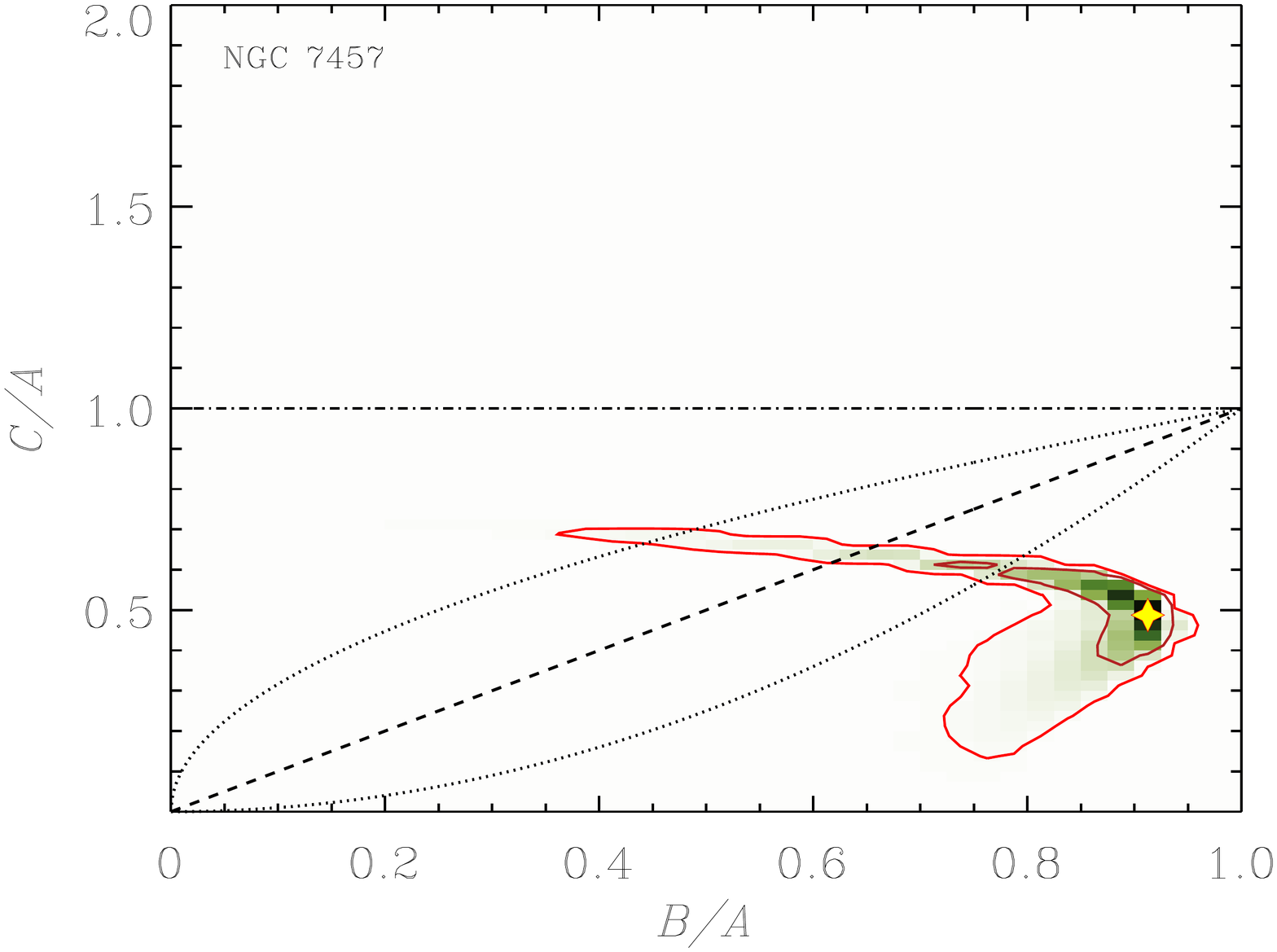}\\
\includegraphics[width=7.7cm, trim=0.5cm 2.4cm 2cm 1.1cm, clip=true]{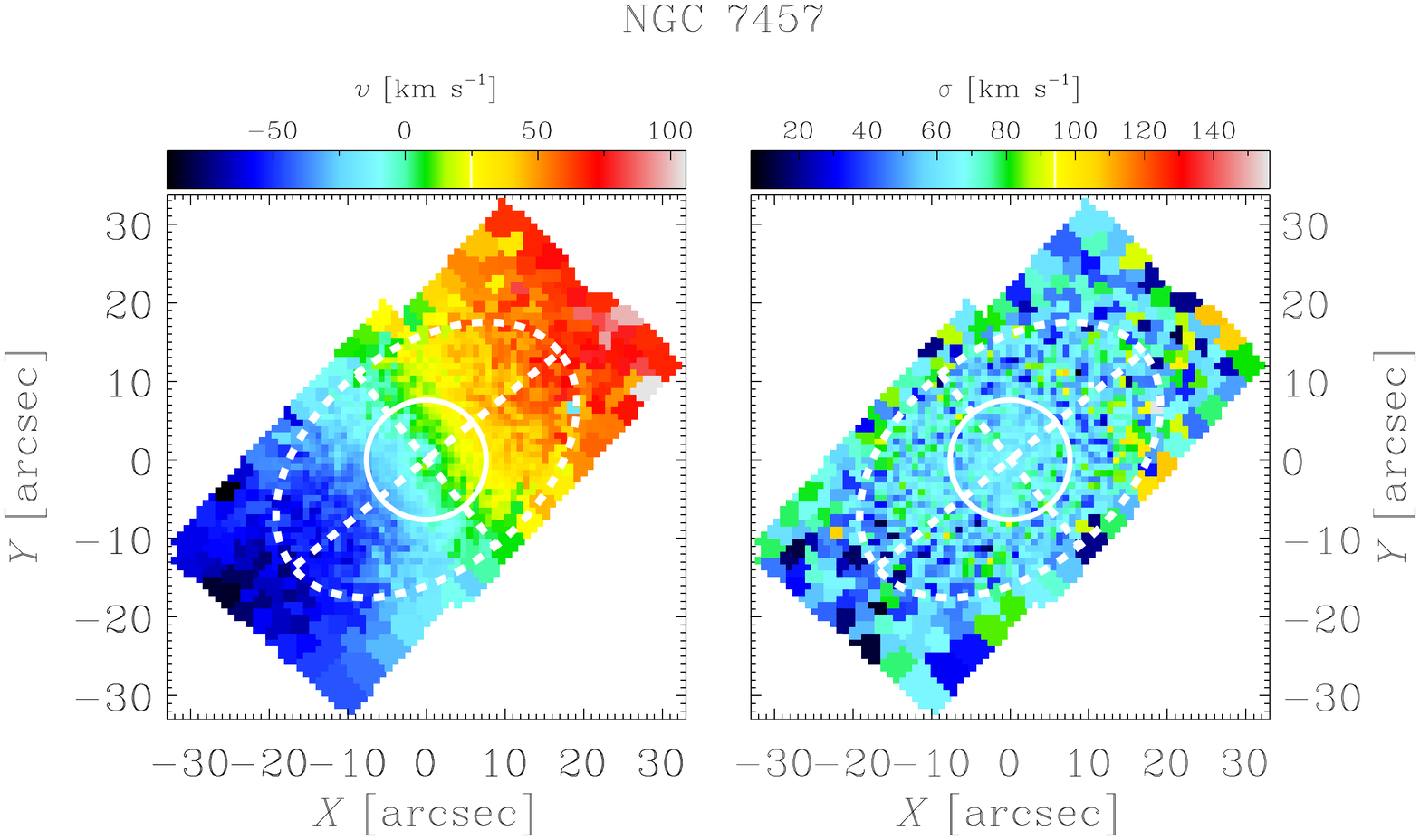}
\includegraphics[width=7.7cm, trim=1.5cm 2.4cm 1cm 0.7cm, clip=true]{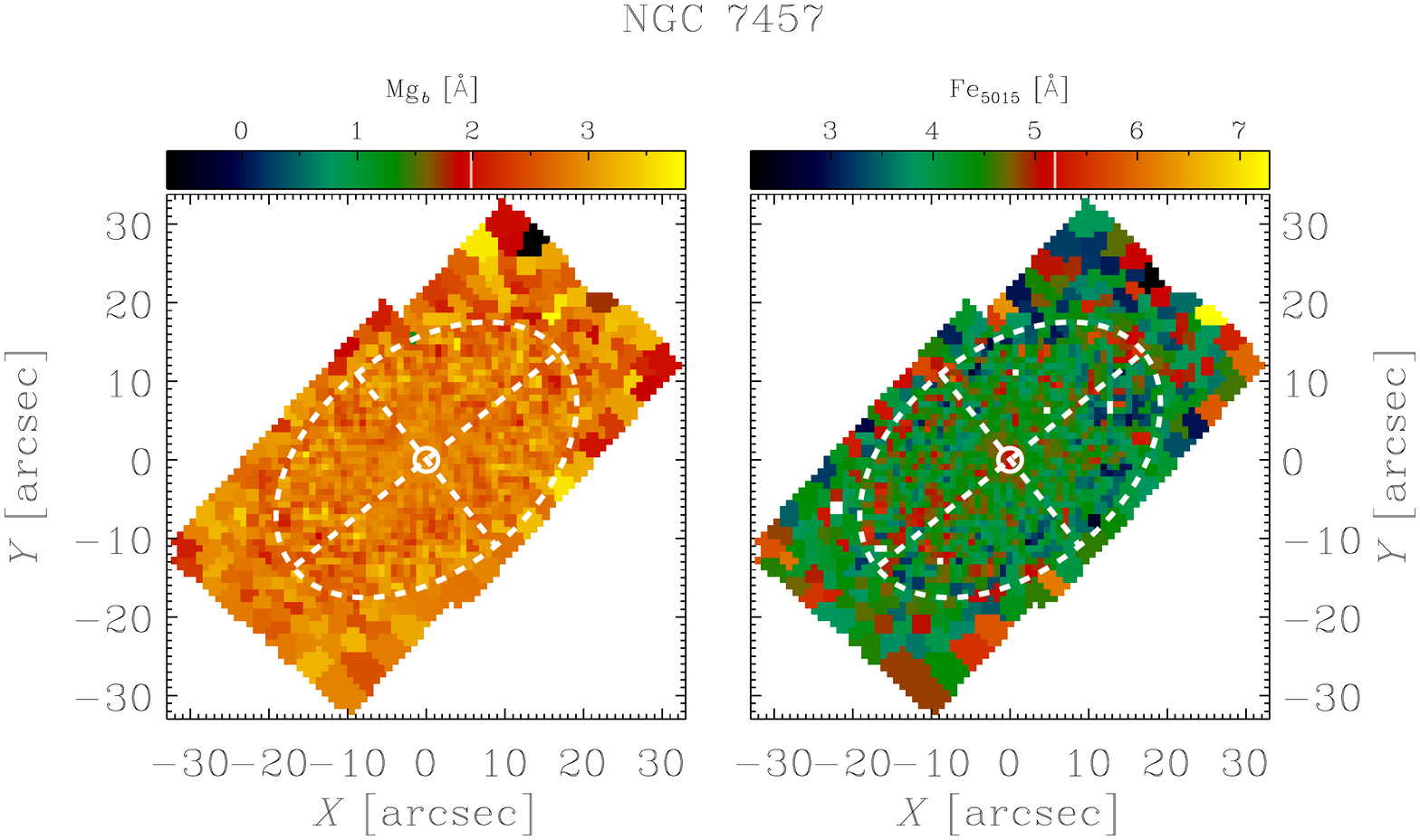}
\caption{continued.}
\end{figure*}

\label{lastpage}
\end{document}